\documentclass[usenatbib]{mnras}

\usepackage{newtxtext,newtxmath}
\usepackage[T1]{fontenc}

\usepackage{graphicx}	
\usepackage{amsmath}
\usepackage{soul,xcolor}

\newcommand{\wi}{\omega_{\rm pi}}

\newcommand{\mime}{m_{\rm i}/m_{\rm e}}
\newcommand{\kB}{k_{\rm B}}
\newcommand{\me}{m_{\rm e}}
\newcommand{\mi}{m_{\rm i}}

\newcommand{\Frad}{\mathcal{F }_{\rm rad}}
\newcommand{\Fw}{\mathcal{F}_{\rm sc}}
\newcommand{\hatFw}{\hat{\mathcal{F}}_{\rm sc}}
\newcommand{\hatFwo}{\hat{\mathcal{F}}_{\rm sc,0}}
\newcommand{\M}{\mathcal{M}}
\newcommand{\bp}{\beta_{+}}

\newcommand{\be}{\beta_{-}}
\newcommand{\bsc}{\beta_{\rm sc}}
\newcommand{\gsc}{\gamma_{\rm sc}}
\newcommand{\ginf}{\gamma_\infty}
\newcommand{\usc}{u_{\rm sc}}
\newcommand{\ue}{u_{-}}
\newcommand{\up}{u_{+}}
\newcommand{\ri}{\mime}
\newcommand{\rif}{\frac{m_{\rm i}}{m_{\rm e}}}

\DeclareMathAlphabet\mathbfcal{OMS}{cmsy}{b}{n}

\setstcolor{red}

\title[Plasma instabilities in relativistic RMS waves]{The role of plasma instabilities in relativistic radiation mediated shocks: stability analysis and particle-in-cell simulations}

\author[A. Vanthieghem et al.]{
A. Vanthieghem,$^{1}$\thanks{E-mail: vanthieg@slac.stanford.edu}
J. F. Mahlmann,$^{2}$
A. Levinson,$^{3}$
A. Philippov,$^{4}$
E. Nakar,$^{3}$
F. Fiuza$^{1}$
\\
$^{1}$High Energy Density Science Division, SLAC National Accelerator Laboratory, Menlo Park, California 94025, USA\\
$^{2}$ Department of Astrophysical Sciences, Peyton Hall, Princeton University, Princeton, NJ 08544, USA\\
$^{3}$ School of Physics and Astronomy, Tel Aviv University, Tel Aviv 69978, Israel\\
$^{4}$ Center for Computational Astrophysics, Flatiron Institute, 162 Fifth Avenue, New York, NY 10010, USA
}

\pubyear{2021}

\begin{document}

\label{firstpage}
\pagerange{\pageref{firstpage}--\pageref{lastpage}}
\maketitle

\begin{abstract}
Relativistic radiation mediated shocks (RRMS) likely form in prodigious cosmic explosions.  The structure and emission of such shocks is regulated by copious production of electron-positron pairs inside the shock transition layer. It has been pointed out recently that substantial abundance of positrons inside the shock leads to a velocity separation of the different plasma constituents, which is expected to induce a rapid growth of plasma instabilities.
In this paper, we study the hierarchy of plasma microinstabilities growing in an electron-ion plasma loaded with pairs and subject to a radiation force. Linear stability analysis indicates that such a system is unstable to the growth of various plasma modes which ultimately become dominated
by a current filamentation instability driven by the relative drift between the ions and the pairs. These results are validated by particle-in-cell simulations that further probe the nonlinear regime of the instabilities, and the pair-ion coupling in the microturbulent electromagnetic field. Based on this analysis, we derive a reduced transport equation for the particles via pitch angle scattering in the microturbulence and demonstrate that it can couple the different species and lead to nonadiabatic compression via a Joule-like heating. 
The heating of the pairs and, conceivably, the formation of nonthermal distributions, arising from the microturbulence, can affect the observed shock breakout signal in ways unaccounted for by current single-fluid models.
\end{abstract}

\begin{keywords}
shock waves -- radiation mechanisms: general -- plasmas -- instabilities -- methods: analytical -- methods: numerical
\end{keywords}

%%%%%%%%%%%%%%%%%%%%%%%%%%%%%%%%%%%%%%%%%%%%%%%%%%

%%%%%%%%%%%%%%%%% BODY OF PAPER %%%%%%%%%%%%%%%%%%

\section{Introduction}
\label{sec:introduction}

Radiation mediated shocks (RMS) are of great interest because they dictate the properties of the early emission seen in a variety of transient 
sources, including various types of supernovae, low luminosity gamma-ray bursts (GRBs),  binary neutron star (BNS) mergers and tidal disruption events (for recent reviews see \citealt{katz2017,levinson2020}).  This early emission, which is released during the 
breakout of the RMS from the opaque envelope enshrouding the central source (or explosion center), carries a wealth of information regarding
the explosion mechanism and the properties of the progenitor, e.g., the radius, density profile, composition and velocity 
profile (in case of BNS mergers), as well as enhanced mass loss episodes that may occur just prior to the 
supernova explosion in certain systems \citep[e.g.,][]{galyam2014,shiode2014}.
A considerable effort has been invested in recent years into the development of dedicated observational programs and new infrastructure that will enable detection of this emission over a wide range of the electromagnetic spectrum (e.g., ZTF, BlackGEM, LSST, ULTRASAT, eROSITA). 
However, the interpretation of such data requires proper modelling of the RMS structure and microphysics, and much
effort has been devoted in the last decade to construct analytical \citep{levinson2008,katz2010,nakar2012,levinson2012,granot2018,lundman2018a} and numerical \citep{budnik2010,beloborodov2017a,ito2018a,ito2020,ito2020b,lundman2021} models 
of Newtonian as well as relativistic RMS. These models have been exploited to compute the structure and spectrum of the breakout emission in different systems. 

Unlike collisionless shocks, in which dissipation is mediated by collective plasma processes on skin depth scales, RMS are mediated by Compton scattering and, if fast enough, electron-positron (e$^\pm$) pair creation. Since the radiation force is felt primarily by the e$^\pm$ pairs, the question arises as to how the ions, which are the primary carriers of the shock energy, are decelerated.  All current RMS models (Newtonian and relativistic) tacitly assume that the multi-species plasma (consisting of photons, ions, electrons and in RRMS also positrons) behaves as a single fluid, which implies infinitely strong coupling between all plasma constituents.

Recently, it has been shown \citep{levinson2020b} that in relativistic and mildly relativistic RMS, in which positrons are abundant, the 
coupling of ions and pairs must involve plasma instabilities, that may actually dominate the shock physics, and may considerably alter the shock structure and emission. This is in contrast with subrelativistic RMS that propagate in purely hydrogen gas, where a tiny charge separation induced by the radiation force gives rise to the generation of an electrostatic field that strongly couples ions and electrons. 
The reason for this important difference is  that the electric field required to decelerate the ions exerts opposite forces on electrons and
positrons that, in turn, leads to separation of the different species already at the onset of the shock transition layer.
The large velocity separation between the electron, positron and ion fluids should lead to generation of plasma turbulence on kinetic scales that can provide the required coupling mechanism. Indeed, inclusion of phenomenological friction forces in the multi-fluid RRMS model indicated that momentum transfer between ions and pairs by anomalous scattering is expected to occur over length scales of hundreds to thousands of skin depths. However, a detailed study of the dominant microinstabilities and associated turbulence is still needed to quantify the degree to and scale over which the different species are coupled.

A mixed ion composition of the upstream flow may complicate the problem further \citep{derishev2018}. The point, again, is that an electrostatic field
cannot decelerate ions with different charge-to-mass ratios at the same rate, so that separation of the different ion beams must ensue. This applies also to Newtonian shocks that propagate in ejecta with mixed ion composition, 
as expected essentially in all supernovae and in BNS mergers.  Because of the large inertia of the ions compared to positrons, the details of the microphysics involved in such shocks might be different than that of a pure-hydrogen, relativistic RMS, but the generation of instabilities and anomalous heating seems unavoidable.  

Generation of plasma turbulence inside the RMS transition layer may  strongly alter the shock breakout signal. Specifically, if  second order Fermi acceleration is effective, then a non-negligible fraction of the shock energy may be tapped to
inject pairs to suprathermal energies, or even to form a population of electrons/positrons with a power law energy distribution. 
This can lead to formation of a hard spectral component that extends well above the spectral energy distribution peak. Moreover, if proton acceleration can also ensue, then RMS can be effective neutrino sources by virtue of the large photo-pion opacity
naturally existing in these shocks. 

In this paper, we study the response of the multi-fluid plasma to the velocity separation between the different species imposed by the radiation force, using a simplified model. In this model the radiation force is replaced by a constant force acting on the leptons (the pair creation process itself is ignored) and the system is taken to be initially uniform within the computation domain. Such conditions may approximately prevail in the flow on scales much smaller than the radiation scales, and are relevant in cases where the induced instabilities grow over timescales much shorter than the deceleration time of the flow.  Our analysis is based on a comparison between analytic and semi-analytic estimates of the growth rates of various plasma modes (\S \ref{sec:plasmainstability}) and particle-in-cell (PIC) simulations (\S \ref{sec:simulations}).  We generally find that the generation of electromagnetic turbulence is mediated by the growth of the filamentation, or Weibel-type, instability \citep[\emph{e.g.}][]{Fried_1959,Achterberg_2007_I} once the velocity separation between ions and leptons roughly exceeds a thermal cutoff obtained by solving the kinetic dispersion relation.  We derive a reduced (semi-)analytical model for the acceleration/deceleration and Joule-type heating of the species in the noninertial microturbulence, accounting for the radiation force and electrostatic field, and show that it can recover the heating and coupling between the different species observed in the PIC simulations (\S \ref{sec:coupling}).
Finally, we discuss the role of an ambient magnetic field in both quenching plasma instabilities and coupling the species (\S \ref{sec:instabilitymagnetized}). Throughout the paper we use Gaussian units with $\kB\,=\,1$. For  the sake of clarity, the three-velocity $\beta$, Lorentz factor $\gamma\,=\,1/\sqrt{1-\beta^2}$ and four-velocity $u\,=\,\beta\gamma$ of the electrons and positrons, with respective subscripts $_-$ and $_+$, are expressed in the instantaneous \emph{ion frame}, except when explicitly stated otherwise. The scattering center frames of the microturbulent field is indicated by subscripts $_{|\rm sc}$.

\begin{figure*}
	\begin{center}
		\includegraphics[width=0.95\textwidth]{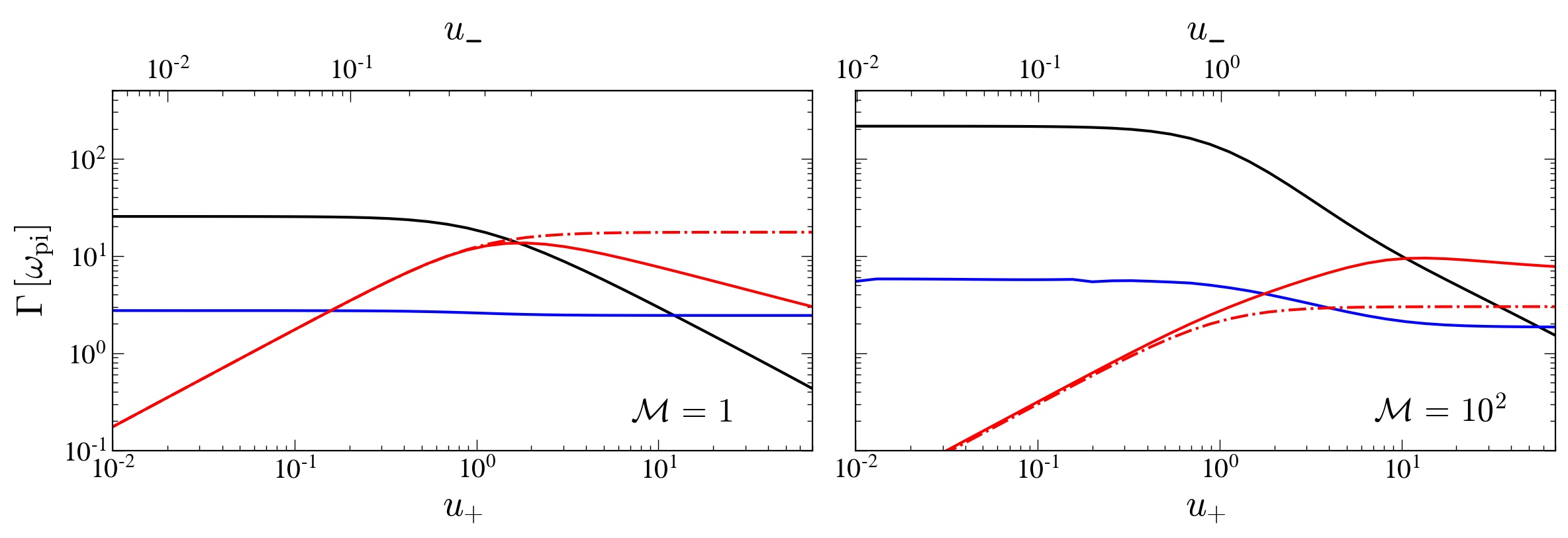}\\
		\caption{Dominant growth rate for the Buneman (blue), two-stream (black) and CFI (red) instabilities for cold plasmas of multiplicities $\mathcal{M}\,=\,1$ (left) and $\mathcal{M}\,=\,10^2$ (right) in terms of the positron four-velocity $\up$. The corresponding four-velocity of the electrons $\ue$ ensuring charge and current neutrality is shown on the superior axis. The dot-dashed line corresponds to the estimate from relation~\eqref{eq:lims_3}. We use a realistic mass ratio $\mime\,=\,1836$.
        }
		\label{fig:disp_long}
	\end{center}
\end{figure*}

\section{Unstable modes in a pair loaded plasma}
\label{sec:plasmainstability}

The construction of a global shock model that can resolve plasma kinetic effects is extremely challenging, perhaps even infeasible. Here, we use a simplified setup retaining the principal features at play in such environments. We consider a system composed of an initially uniform electron-ion background plasma loaded with pairs of multiplicity $\mathcal{M}$ relative to the proper ion background density $n_{\rm i}$. Normalized electron and positron densities thus read $\left(\mathcal{M} + 1 \right)$ and $\mathcal{M}$, respectively. The effect of the photon field reduces to an external homogeneous and constant radiative force $\mathcal{F}_{\rm rad}$ acting on the leptons. In the following, $\mathcal{F}_{\rm rad}$ is expressed in units of $m_{\rm i} \wi c$, where $m_{\rm i}$ is the ion mass, $\wi = \sqrt{4 \pi n_{\rm i} Z^2 e^2/m_{\rm i}}$ the ion plasma frequency, $e$ the electron charge, and $Z$ the ion charge ($Z\,=\,1$ here).  
In real RMS the value of $\mathcal{F}_{\rm rad}$ changes throughout the shock transition layer, and scales roughly 
as $(\ginf/\gamma)^2$, where $\ginf$ is the upstream Lorentz factor and $\gamma$ the local Lorentz factor.  
For $\ginf=10$ the typical value of $\Frad$ well inside the shock, where the pair multiplicity approaches the mass ratio, is around a few times $10^{-8}$ in infinite RRMS, and higher by a factor of up to 10 during the breakout phase, $\mathcal{F}_{\rm rad}\sim10^{-7}$. 

Our approach thus consists in computing a stability analysis around the unperturbed trajectory of each fluid. In such a reduced system and in the nonrelativistic regime to order $\tilde{w}_e/m_i$, the unperturbed electric field reads $E^0\,\simeq\,\Frad/(1 + 2 \M) \left[1 - \cos\left(  \omega_0 t \right) \right]$, where $\omega_0/\wi\,=\,\sqrt{\left(1 + 2 \M\right) \mi/\tilde{w}_e}$, and $\tilde{w}_e$ is the electron enthalpy normalized to its proper density. The radiation force will lead to a relative drift between the leptons and the ions and the electrostatic field will then give rise to relative drifts between the electrons and the positrons. Because these relative drift velocities are subrelativistic to mildly relativistic at the onset of plasma instabilities, as we discuss below in more detail, we consider the nonrelativistic limit a good approximation on the growth time of the instability.

In this section, we consider the development of plasma instabilities associated with the relative drift between the different plasma species induced by the radiation force and the electrostatic field. We consider that the most unstable modes grow much faster than the deceleration time of the fluid, which is in reality imposed  by the radiation intensity inside the shock precursor. This allows us to neglect the effect of the radiative force and electrostatic field during the instability growth time. This approximation is verified in \S \ref{sec:simulations} during the instability growth time using fully kinetic PIC simulations. We carry out a linear stability analysis of the multi-fluid system, consisting of an electron-ion plasma loaded with pairs of multiplicity $\M$.
We consider the cold regime first, and then extend the analysis to a pair distribution with finite temperature
by making use of a kinetic description to gauge thermal effects for various multiplicities.

\subsection{Cold plasma limit}
\label{sec:cold}

In the following, we derive the longitudinal and transverse linear unstable modes in the cold limit and study their dependence on the relative drift velocity between the species, while imposing charge and current neutrality. These longitudinal and transverse modes, characterized by a dominant wavevector respectively aligned or transverse to the flow, are obtained by solving the 3-species cold relativistic dispersion relation. The general dispersion relation for the purely longitudinal or transverse modes is shown in Eq.~\eqref{eq:disp_rel_kin_cold} and Eq.~\eqref{eq:disp_rel_kin} for which a cold distribution is assumed (see~\citealt{Melrose_1986}).

\emph{Longitudinal modes} -- Purely electrostatic modes are usually the fastest growing modes in mildly to ultra relativistic cold beam-plasma systems. However, these tend to saturate rapidly and at a relatively low amplitude. Figure~\ref{fig:disp_long} shows the dominant electrostatic modes (black and blue curves) for two different multiplicities $\mathcal{M}\,=\,1$ (left) and $\mathcal{M}\,=\,10^2$ (right) as a function of the relative drift velocity between the leptons and the ions. The black and blue curves correspond to electrostatic electron-positron two-stream and pair-ion Buneman branches of unstable modes, respectively. We observe that for cold plasmas with non- to mildly relativistic relative drift speeds, the two-stream instability is the fastest growing mode. 

\emph{Transverse modes} -- The development of transverse modes, namely the current filamentation instability (CFI) \citep{Fried_1959}, can play an important role in shaping the long-term momentum transfer between the pairs and the ion distributions. The magnetic field perturbations seeded by the CFI are of magnetic nature  $\delta \mathbf{B}^2 - \delta \mathbf{E}^2 > 0$ and the growing  magnetic field is transverse to the flow.
In its full form, the dispersion relation reduces to a polynomial of order 3 in $\omega^2$ obtained from Eq.~\eqref{eq:disp_rel_kin_cold} for a cold distribution.
In the cold plasma approximation, estimates of the maximum growth rate are obtained by taking the large $k$ limit. Even though the dominant branch of the CFI is unique, one can derive two limits that respectively correspond to the pair-ion CFI and electron-positron CFI:
\begin{align}
{\rm e^\pm-i} : \qquad \Gamma\, &= |\bp| \wi \,, \label{eq:lims_1}\\
{\rm e^+-e^-} : \qquad \Gamma\,&= |\bp| \sqrt{ \frac{ \M\,\ri  }{\left( 1 + \mathcal{M} \right) \left( 1 + 2 \mathcal{M} \right) } } \,\, \wi \label{eq:lims_3}\,,
\end{align}
where $\me$ is the electron mass and $\bp$ is the positron drift velocity in the ion frame.  Eq.~\eqref{eq:lims_1} is derived in the limit of infinite multiplicity in which the relative drift between electrons and positrons goes to zero so that electron-positron CFI and two-stream are suppressed. This limit remains valid when the thermal spread of the pairs becomes comparable to the relative drift between electrons and positrons. Note that for infinite multiplicity, the relative drift between electrons and positrons goes to zero, and both longitudinal and transverse electron-positron streaming instabilities are suppressed. Eq.~\eqref{eq:lims_3} is valid for cold plasmas and finite multiplicity. In this limit, the electron-positron CFI is always faster than the pair-ion CFI, however,  the growth rate of the former can be strongly suppressed by thermal effects as discussed below.

The comparison between the growth rates of the transverse CFI mode and the two longitudinal modes is shown in Fig.~\ref{fig:disp_long}. We observe that in the cold limit the CFI modes (red) are subdominant in the nonrelativistic regime.

\subsection{Temperature and multiplicity dependence}
\label{sec:kinetic}

\begin{figure}
	\begin{center}
		\includegraphics[width=1.\columnwidth]{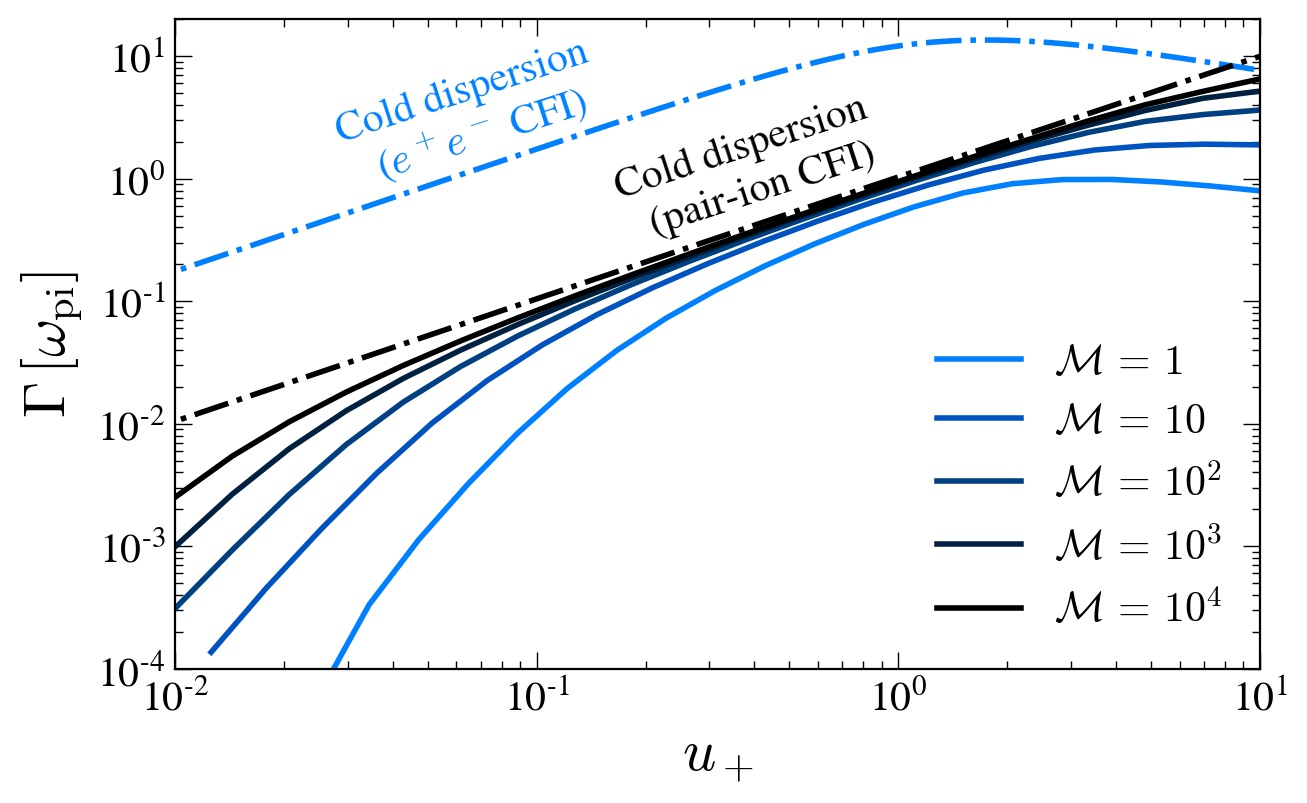}\\
		\caption{Analytic growth rate versus the positron four-velocity $\up$ in the ion frame, for relativistic pair plasmas of temperature $ T_\pm\,=\,\me c^2$ and multiplicities $\M\,=\,1, \, 10,\, 10^2,\, 10^3$ and  $10^4$ (from light blue to black). The dotted-dashed lines correspond to the maximum growth rate in the cold limit, for $\M\,=\,1$ (blue) and $\M\,\gg\,1$ (black). A realistic mass ratio, $\mime=1836$, has been adopted in all cases shown.
        }
		\label{fig:G_u_th}
	\end{center}
\end{figure}

In realistic RRMS conditions, the temperature in the immediate downstream is regulated by pair creation at a value of about $m_ec^2/3\sim 200$ keV \citep{katz2010,granot2018,levinson2020}.  Inside the shock the 
temperature is somewhat higher, reaching values of about $0.4\gamma_s m_ec^2 \sim$ a few MeV, where $\gamma_{\rm sh}$ is the shock Lorentz factor \citep{granot2018,ito2020}.
While the cold approximation remains valid for the ions, the thermal spread of the leptons must be included in the stability analysis in order
to properly compute the development of electron-positron and pair-ion instabilities. In this regime, electron-positron two-stream modes are strongly damped as the relative drift between leptons is comparable or smaller than the thermal spread. This is especially true in the regime of high multiplicity where the relative drift between leptons satisfies $|\Delta \beta | \simeq |\bp|/\mathcal{M} \ll 1 $.
Hereafter, we thus focus on the generation of transverse modes. To gain some insight, we first approximate the pairs by a Maxwellian
distribution, assuming that the relative drift between warm leptons and cold ions remains subrelativistic to mildly relativistic. The unstable transverse electromagnetic modes are obtained by solving the corresponding kinetic dispersion relation as described in~\cite{Silva_2002}.
The complete derivation is presented in Appendix \ref{sec:appA}. Here, we merely present the main results. 

While the electron-positron CFI is the dominant transverse mode for any value of the relative drift in the cold case, it is strongly damped by thermal effects once the relative drift between electrons and positrons becomes smaller than the thermal spread (similarly to the case of electrostatic modes). Quantitatively, for a leptonic temperature $T_\pm$ these transverse modes are negligible when
\begin{equation}
\bp^2\,\lesssim\,\frac{T_\pm}{\me c^2} \left( \M + 1 \right)^2\, \label{eq:CFI_pm}
\end{equation}
and thus are not expected to be important even at low multiplicities ($\M\,\sim\,1$). On the other hand, the branch that corresponds to the pair-ion CFI in Eq.~\eqref{eq:lims_1} gives a good approximation for the full solution of Eq.~\eqref{eq:disp_Max}, provided the pair temperature is sufficiently low. To gauge the range of validity of Eq.~\eqref{eq:lims_1}, thermal effects can be probed by perturbing the full solution around this approximation -- i.e., $\Gamma\,\simeq\,|\bp|\,\wi \left(1 + \delta \Gamma \right)$. We estimate the wavenumber $k_{\rm max}$ of the dominant mode as of being of the order of the maximum unstable mode fixed by the thermal cutoff at large $k$.
Following this argument, the typical wavenumber reads 
\begin{equation}
k_{\rm max}\,\sim\,|\bp| \sqrt{\frac{\mathcal{M}\,\mi c^2}{T_\pm}}\,\wi/c\,,
\end{equation}
up to a correction factor of order unity. Substituting the above result into Eq.~\eqref{eq:disp_Max}, we obtain, to leading order in $\delta\Gamma$, the domain of validity of Eq.~\eqref{eq:lims_1}. The growth rate is considerably quenched by thermal effects when 
\begin{equation}
\bp^2\,\lesssim\, \frac{T_\pm}{\me c^2} \sqrt{\frac{1}{\M}\frac{\me}{\mi}}\,.\label{eq:CFI_pi}
\end{equation}
Interestingly, we see that even at mildly relativistic pair temperatures ($T_\pm\sim \me c^2$) and low multiplicities ($\M \sim 1$), a velocity of the order of $\bp\,\sim\,0.1$ is sufficient for the pair-ion CFI growth rate to approach its approximate value given by the cold limit of Eq.~\ref{eq:lims_1}, and that higher multiplicities are even more favorable.

Because relativistic effects were not taken into account in the previous derivations, we verify the analytical results by integrating numerically Eq.~\eqref{eq:disp_rel_kin} in the relativistic regime. The electron and positron distributions are approximated by relativistic drifting Jüttner-Synge distributions:
\begin{equation}
    f_\pm(\mathbf{p}) \,=\,\frac{\gamma_\pm n_\pm m_e c^2 }{4 \pi  T_\pm K_2(m_e c^2/T_\pm)} e^{-\frac{\gamma_\pm m_e c^2}{T_\pm} \left( \sqrt{1+\mathbf{p}^2} - \beta_\pm p_x \right)} \,,
\end{equation}  
where $\beta_\pm$ is the electron/positron drift speed, $\gamma_\pm\,=\,(1-\beta_\pm^2)^{-1/2}$ and $K_2$ is the modified Bessel function of the second kind. The ions are assumed to have a cold distribution. For a covariant approach to the kinetic dispersion relation, see~\cite{Achterberg_2007_I}. The comparison between the cold and warm growth rates, obtained from the numerical integration of the dispersion equation, as a function of the relative drift, is shown in Fig.~\ref{fig:G_u_th}. For relativistic temperatures, the electron-positron CFI is suppressed, as expected from Eq.~\eqref{eq:CFI_pm}. One also recovers the dominant branch of pair-ion CFI and the thermal cutoff $\beta_{\rm cutoff}\propto\M^{-1/4}$ consistent with our kinetic estimate in Eq.~\eqref{eq:CFI_pi}.
From these considerations we deduce that transverse modes can grow substantially when the relative drift between ions and pairs becomes of the order of $\up\,\sim\,0.1$.

\subsection{The scattering center frame and saturation level}
\label{sec:saturation}

\begin{figure}
	\begin{center}
		\includegraphics[width=1.\columnwidth]{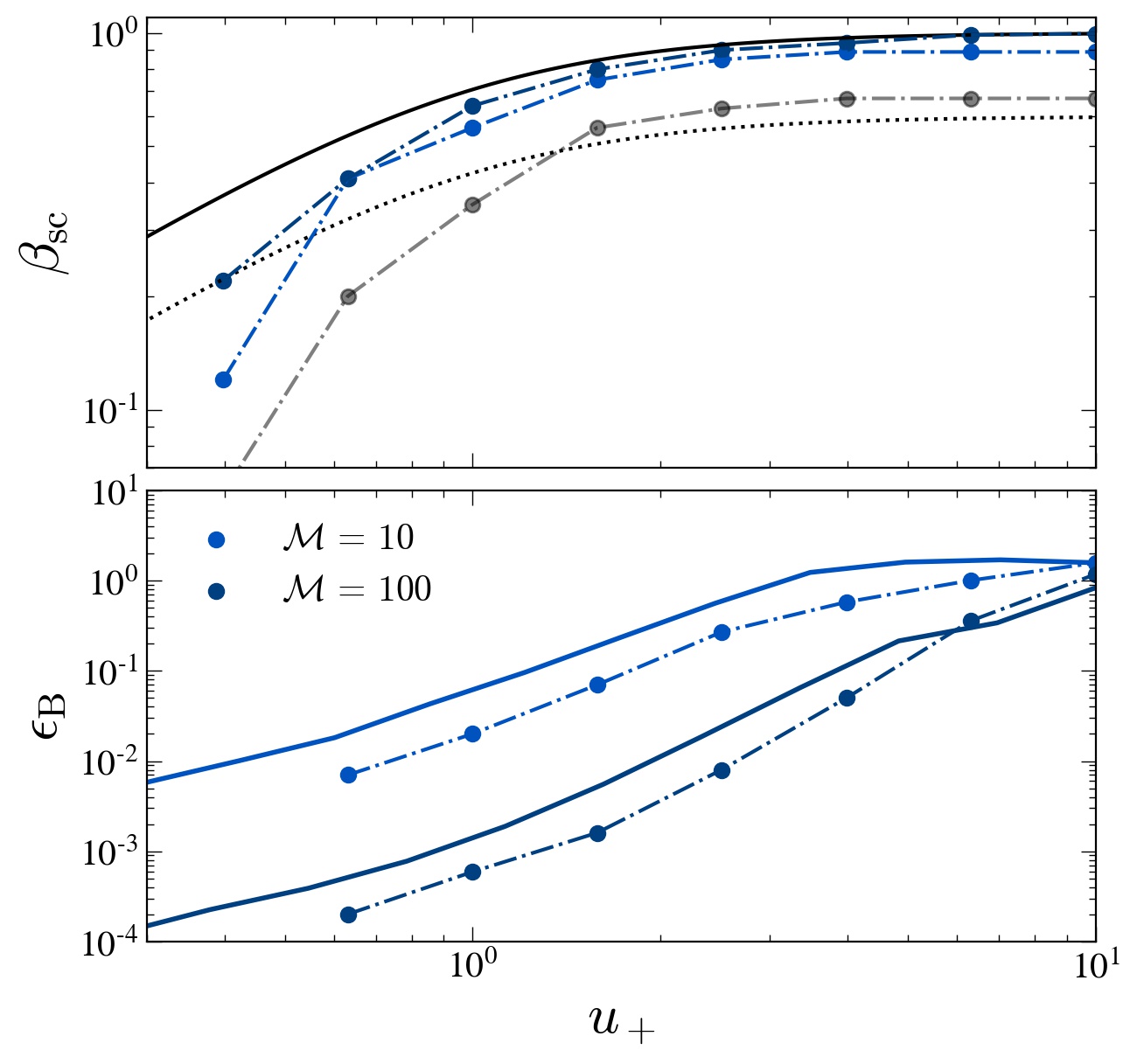}\\
		\caption{Top: velocity of the scattering center as measured in the ion frame. The solid line corresponds to $\bp$ and the dotted black line to the low multiplicity estimate of the scattering center frame $\beta_{\rm sc}\,\simeq\,0.6\bp$ from Eq.~\eqref{eq:sc_frame_est}. Bottom: magnetization in units of $4 \pi n_{+} m_{\rm e } c^2$ at saturation of CFI obtained from magnetic trapping criterion for various multiplicities in terms of the relative drift between the positrons and ions. The plasma temperature is set to $T_\pm \,=\,\me c^2$. The points joined by dashed-dotted lines correspond to the measurements from 1D PIC simulations. Solid lines correspond to the saturation level obtained by solving the fully kinetic relativistic dispersion relation described in \S \ref{sec:kinetic} and indicate overall good agreement with the simulation results. The gray dashed-dotted line corresponds to the case $\M\,=\,1$.}
		\label{fig:sat}
	\end{center}
\end{figure}

We have seen that pair-ion CFI modes will grow and become dominant for relativistic pair temperatures and subrelativistic to mildly relativistic drift velocities between the pairs and the ions. The growth of the CFI will give rise to magnetic turbulence that can scatter the particles. The scattering efficiency depends on the magnetic field structure and saturation level of the CFI. The kinetic energy of the species in the scattering center frame is crucial to derive the saturation criterion of the instability. The scattering center frame corresponds to the frame in which the boosted transverse component of the electric field is negligible compared to the amplitude of the magnetic field -- \emph{i.e.} $\delta \mathbf{B}_\perp^2 \gg \delta \mathbf{E}_\perp^2 $. Following the approach of~\cite{Pelletier_2019}, the 3-velocity of the scattering center frame in a three-fluid plasma, consisting of cold ions and pairs at finite temperature, satisfies
\begin{equation}
    \frac{\wi^2}{(\omega/kc)^2} \beta_{\rm i|sc} + \sum_\pm \frac{\omega_{\pm}^2}{\tilde{w}_\pm} \beta_{\pm \rm |sc} \frac{1 - c^2_{\pm,\rm eff}}{(\omega/kc)^2 - c^2_{\pm, \rm eff}}\,=\,0\,, \label{eq:sc_frame}
\end{equation}
where $c^2_{\pm,\rm eff}\,=\,c_{\rm s}^2/\gamma_\pm^2(1-c_{\rm s}^2\beta_\pm^2)$ with $c_{\rm s\pm}\,=\,\sqrt{\Gamma_{\rm ad} P_\pm/w_\pm}$ being the isentropic sound speed, $\Gamma_{\rm ad}$ the adiabatic index of the pair plasma, and $\beta_{\alpha\rm |sc}$ the velocity of species $\alpha$ relative to the scattering center frame. The above Eq.~\eqref{eq:sc_frame} is then solved for the scattering center velocity $\beta_{\rm sc}$ in the ion frame writing $\beta_{\alpha\rm |sc}\,=\,(\beta_{\alpha} - \beta_{\rm sc})/(1 - \beta_{\alpha}  \beta_{\rm sc} )$. While exact solutions are straightforward to derive, simple analytical expressions are hampered by the mode dependence that we observe in our simulations, as well as by the relativistic formulation. Thus, for the sake of conciseness, we derive here the simple mode-independent relation for a hot pair plasma ($T_\pm\gtrsim\,m_e c^2$) drifting at subrelativistic speed in the ion frame, for which $c^2_{\pm,\rm eff}\simeq 1/2\gamma_\pm^2 \ll 1$, assuming $|\omega/kc|^2\,\gg\,c^2_{\pm,\rm eff}$. Equation~\eqref{eq:sc_frame} then reduces to:
\begin{equation}
    \beta_{\rm i|sc}\frac{\me}{\mi} + \M \beta_{+\rm|sc} \frac{1 - \Gamma^{-1}_{\rm ad}}{T_{+}/\me c^2} + (\M+1) \beta_{-\rm|sc} \frac{1 - \Gamma^{-1}_{\rm ad}}{T_{-}/\me c^2} \,=\,0 \,.
\end{equation}
To leading order in $\me/\mi$, using charge neutrality with $\M \beta_+ = (\M+1)\beta_-$, and for $\gamma_+ \simeq 1$,
the solution for the velocity of the scattering center frame reads:
\begin{equation}
    \beta_{\rm sc} \,\simeq\, \frac{ \M}{T_+/\left( T_+ + T_- \right) +  \M} \bp\,.
    \label{eq:sc_frame_est}
\end{equation}
We have checked that this relation provides a good estimate of the exact solution even in the relativistic regime. In the low multiplicity regime ($\M\,=\,1$), we have $\beta_{\rm sc}\,\sim\,0.6\,\bp$. When $\M\,\gg\,1$, the scattering center drifts along with the bulk of positrons and electrons and their temperatures are almost equal. 

The top panel of Fig.~\ref{fig:sat} shows the velocity of the scattering center frame as obtained from a series of 1D PIC simulations performed with the electromagnetic, relativistic code OSIRIS ~\citep{Fonseca_OSIRIS,Fonseca_2008} for multiplicities $\M\,=\,1,\,10,\,100$. These 1D PIC simulations resolve the direction transverse to the flow. We initialize velocities and densities for various relative drifts between positrons and ions, imposing charge and current neutrality.  

In the high multiplicity regime, ions are clearly responsible for the saturation of the CFI as their kinetic energy in the scattering center frame is the source of the instability. An estimate of the saturation level can be obtained from the magnetic trapping condition in the filaments \citep[\emph{e.g.}][]{Yang_1994,Lyubarsky_2006}. The saturation of the CFI occurs when the bounce frequency of the ions inside a filament becomes comparable to the growth rate of the instability, \emph{i.e.} $\omega_{\rm B}\,\sim\,\Gamma_{\rm CFI}$, with $\omega_{\rm B}$ the ion bounce frequency satisfying $\omega_{\rm B}^2\,=\, e \delta B_{\rm |sc} \,k_{\rm B}\beta_{\rm i|sc}/\gamma_{\rm i|sc}\mi c$, where $\delta B_{\rm |sc} $ and $\beta_{\rm i|sc}$ the respective magnetic field amplitude and ion velocity in the scattering center frame.
A comparison of the predicted saturation and magnetization level, in units of $4 \pi n_{+} m_{\rm e} c^2$, with 1D PIC simulations (dot-dashed lines) is shown in the bottom panel of Fig.~\ref{fig:sat}.

\begin{figure*}
	\begin{center}
		\includegraphics[width=1.\textwidth]{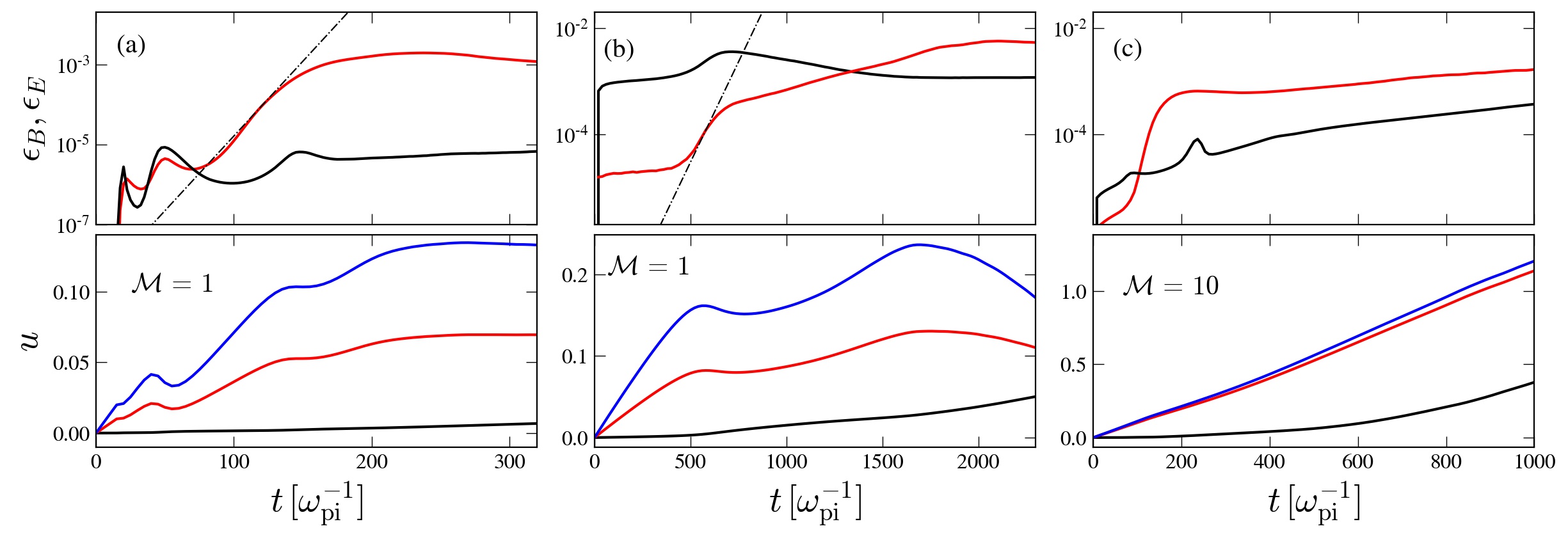}\\
		\caption{Magnetization and drift velocity of the species as obtained from PIC simulations. The three columns correspond to different initial pair temperature and/or multiplicity: (a) cold $T_\pm\,=\,10^{-8} \me c^2$, $\M\,=\,1$, $\Frad\,=\,10^{-5}$, and $\mime\,=\,100$; (b) mildly relativistic $T_\pm\,=\, \me c^2$, $\M\,=\,1$, $\Frad\,=\,10^{-5}$, and $\mime\,=\,100$; (c) mildly relativistic $T_\pm\,=\,0.1\me c^2$, $\M\,=\,10$, $\Frad\,=\,9.5 \times 10^{-5}$, and $\mime\,=\,25$.  Top panel: magnetic field (red) and electric (black) energy density in units of $4 \pi n_{+} m_{\rm e} c^2$. In the top middle, the solid line correspond to the magnetic spectral energy density of the modes contained in the envelope of the CFI linear unstable modes. Bottom panel: longitudinal four velocity of the positrons (blue), electrons (red) and ions (black) in the initial frame of the ions. 
        }
		\label{fig:M1_PIC}
	\end{center}
\end{figure*}

\section{PIC simulations and nonlinear evolution}
\label{sec:simulations}

\begin{figure*}
	\begin{center}
		\includegraphics[width=0.84\textwidth]{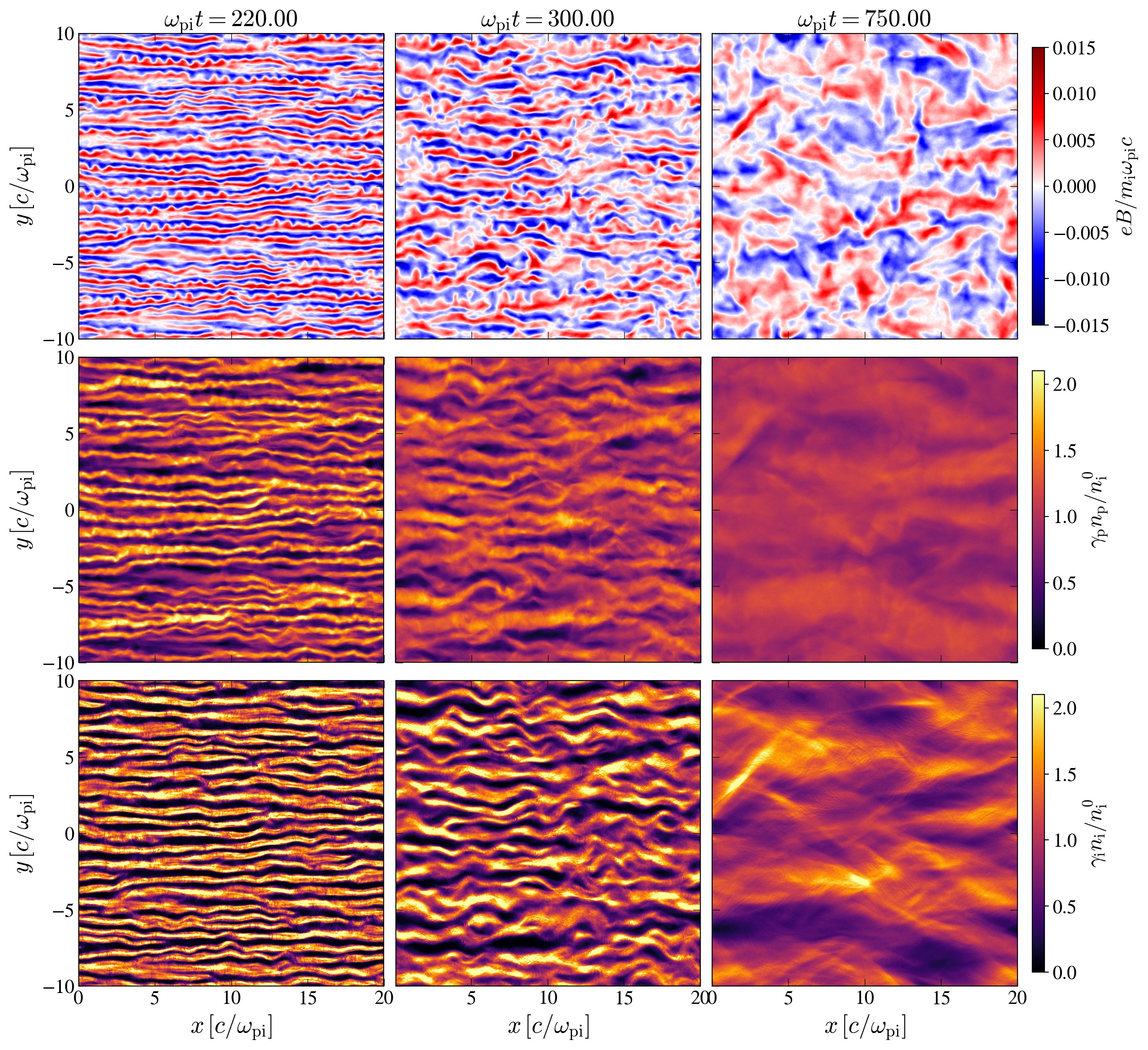}\\
		\caption{Transverse magnetic field (top) and apparent density of the positrons,  $\gamma_{\rm p} n_{\rm p}$ (middle), and ions, $\gamma_{\rm i} n_{\rm i}$ (bottom), at three different stages of the current filamentation instability observed in the PIC simulation with an initial cold temperature $T_\pm\,=\,10^{-8} \me c^2$, multiplicity $\M\,=\,1$, and radiation force $\Frad\,=\,10^{-5}$. From left to right: $\wi t = 220, 300, 750$. At $\wi t = 220$, positronic filaments become kink unstable, leading to a longitudinal modulation of the filaments with a typical wavenumber $\wi k_\parallel/c \simeq 12.5$. At later times, the electron-ion current filaments develop larger scales kink modes with $\wi k_\parallel/c\simeq 1.5$ and eventually lead to filament disruption and onset of the microturbulence.
        }
		\label{fig:cold_kink}
	\end{center}
\end{figure*}

In the preceding section we studied the linear stability of a pair-loaded plasma subject to an external force acting on the leptons. We verified that instabilities develop over scales vastly shorter than the radiation scales and are, therefore, expected to establish coupling between the various plasma constituents.  However, the linear analysis cannot account for the influence of the instability on the thermodynamic state of the plasma. Of utmost interest to astrophysical applications is the saturation level of the instability and its effect on the spectrum of the pairs inside the shock and the resultant emission.
Therefore, while benchmarking our linear results, we now focus on subsequent effects such as saturation of the electromagnetic field growth, subsequent instabilities that may arise in the nonlinear phase, and the transfer of momentum between pairs accelerated by the radiative force and ions.  In the following, we describe the self-consistent plasma dynamics obtained from 2D PIC simulations 
and compare the results with our analytical calculations. 

\subsection{Low multiplicity ($\M\,\simeq\,1$)} 
\label{sec:LowM}

We first investigate the configuration in which the density of pairs is comparable to the density of the ions. In that case, the relative drift between the electrons and positrons in the ion frame is of the order of the positron drift speed ($\Delta \beta\,\simeq\,\bp/2$ for $\M=1$). To characterize the dominant unstable modes, we consider two different initial temperatures for the pair plasma: $T_\pm/m_{\rm e} c^2\,=\, (10^{-8},\,1)$. In the simulation setup, all species (electrons, positrons, and ions) are initialized at rest (no drift velocity). The electrons and positrons are then subject to a constant radiative force. We use periodic boundary conditions for both fields and particles with a mesh size $\wi \Delta x/c\,=\,0.2/\sqrt{\M \ri}$ and time step $c \Delta t\,=\,0.5 \Delta x$ with cubic interpolation. Each cell contains 500 macroparticle per species . We use a reduced mass ratio $\mime\,=\,100$, which allows us to capture a sufficiently large scale separation between leptons and ions at manageable computational expense. Instabilities are purely seeded by thermal noise.
Figure~\ref{fig:M1_PIC} illustrates the temporal evolution of the magnetic and electric field energy densities, and the relative drift between species for increasing temperatures. In all cases, the long-term dynamics is dominated by transverse modes --- \emph{i.e.} $\delta \mathbf{B}_\perp^2 - \delta \mathbf{E}_\perp^2 > 0$, where $\delta \mathbf{B}_\perp$ is the transverse, out-of-plane, component of the magnetic field --- excited by the growing relative drift between species.

\subsubsection{Cold pair plasma ($ T \ll \me c^2$)}
\label{sec:LowT}

In the limit of a cold plasma $ T_\pm \ll \me c^2$, the electrostatic and electromagnetic modes discussed in \S \ref{sec:cold} successively grow and quench the phase space anisotropies that drive them.
The first electrostatic mode to grow and saturate around $\wi t\,=\,22$ corresponds to the electrostatic two-stream instability between electrons and positrons. After its saturation, and as the relative drift between ions and pairs increases to about $\Delta \beta\,\simeq\,0.04$, the plasma becomes unstable to Buneman modes around $\wi t\,=\,50$. When these electrostatic modes saturate, and as the electrostatic field decays, the coupling between pairs and ions declines and the pairs are then again efficiently accelerated by $\Frad$. Any resurgence of electrostatic modes is then prevented by the strong pair heating up to $ T_\pm\,\simeq\,10^{-4} \me c^2$ at the saturation of the Buneman instability. 

The transition from electrostatic to electromagnetic mode dominance is clearly observed at $\wi t\,\simeq\,70$, at which time $\epsilon_E\,<\,\epsilon_B$. By that stage, the saturation of the two electrostatic modes efficiently heated the electrons and positrons to temperatures of $T_{\rm -} \simeq 10^{-3} \me c^2$ and $ T_{\rm +} \simeq 1.5\times10^{-3} \me c^2$, respectively. As the drift velocity of the pairs increases due to lack of sufficiently strong coupling to the ions, the system transitions to a magnetic configuration dominated by the CFI. From the magnetic energy density evolution in Fourier space, we extract the approximate maximum growth rate, $\Gamma_{\rm PIC}\,\simeq\,4.7\times10^{-2}\,\wi$, at a transverse wavenumber $ k \,\simeq\,9\,\wi/c$. 

Using the kinetic description outlined in Sec.~\ref{sec:kinetic}, we compare the growth rate with analytical estimates and derive a maximum value of $\Gamma_{\rm th}\,=\,4.3\times10^{-2}\,\wi$ at $k = 8.2\,\wi/c$ for the drift velocity $\bp\,\simeq\,0.07$ around $\wi t\,=\,100$, in good agreement with our PIC simulations. Since the pair heating associated with the early electrostatic instabilities quenches the electron-positron CFI, the analytical estimate in Eq.~\eqref{eq:lims_1} for the ion-pair CFI approximates well the growth rate $\Gamma_{\rm th}\,=\,(5-7)\times10^{-2}\,\wi$.

\subsubsection{Relativisticaly hot pair plasma ($T \sim \me c^2$)}
\label{sec:LowT}

The transfer of momentum between the photons and pair distributions in the shock precursor naturally leads to mildly relativistic pair temperatures, $T_\pm \simeq \me c^2$. As discussed in Sec.~\ref{sec:kinetic}, such thermal effects strongly suppress the growth of both longitudinal (electrostatic) and transverse (electromagnetic) modes associated with relative electron-positron drifts. To assess the stability of a system comprised of cold ions and a warm pair plasma, we now consider a warm initial pair temperature $T_\pm=\, \me c^2$. 
Simulations indicate that when the positron drift velocity exceeds  $\bp\,\sim\,0.1$, the system becomes unstable to both electrostatic and electromagnetic modes at time $\wi t\,\simeq\,500$ in Fig.~\ref{fig:M1_PIC}(b). Analysis of the evolution of the longitudinal electric field and transverse magnetic field in Fourier space confirms the growth of the short-scale longitudinal Buneman mode and the transverse pair-ion CFI mode. We extract the dominant growth rate of the pair-ion CFI in Fourier space and obtain $\Gamma_{\rm PIC}\,\simeq\,8.7\times10^{-3}\wi$ with a dominant wavenumber $k\,=\,1.3\,\wi/c$. At that stage, the temperature and drift velocities respectively read $T_\pm= \me c^2$ and $\bp\,=\,0.16$. 
The numerically solved full dispersion relation leads to a maximum $\Gamma_{\rm th}\,=\,8.7\times10^{-3}\,\wi$ at $k\,=\,1.1\,\wi/c$ for $T_\pm\,=\,\me c^2$, in good agreement with PIC results. We observe that the long-term evolution of the system continues to be dominated by the CFI mode, which is seen between $\wi t \,\simeq\,[560,1700]$ in Fig.~\ref{fig:M1_PIC}(b).
We ran additional simulations for colder temperatures ($T_\pm\sim\,0.1\me c^2$) and found that the system exhibits dynamics similar to the case $T_\pm\simeq\,\me c^2$. 

It is important to note that the presence of the external force is critical to sustain the drift anisotropy between the pairs and the ions and thus enable the development of the dominant CFI modes. We have checked that in the absence of external radiation force, and for the same initial conditions at the onset of the instability ($\bp = 0.16$) as described in the previous paragraph, the pair plasma slowdown due to the electrostatic field produced by the Buneman instability is sufficient to quench the CFI and the resulting amplitude of the microturbulence is significantly lower. 

The final nonlinear stage of the CFI corresponds to merging and disruption of filaments, setting the magnetization level in the shock precursor and sustaining the migration to larger scale modes. We observe that the disruption of the current filaments occurs via the development of the kink instability \citep{Ruyer_2018, Vanthieghem_2018} for both cold and warm pair plasma cases. This is illustrated in Figure~\ref{fig:cold_kink}, which shows the evolution of the magnetic field, positron, and ion density profiles at different times, $\wi t\,=\,220,\,300,\,750$, for a cold plasma of multiplicity $\M\,=\,1$. We observe that after saturation of the CFI, longitudinal kink-type modulations of the positron filaments develop at small scales. The kink instability grows with a typical longitudinal wavenumber $ k \simeq 12.5\,\wi/c$. Around $\wi t \,=\, 300$, the electron-ion current filaments develop longitudinal modes at the ion skin depth scales $k\simeq 1.5\,\wi/c$. For hot positrons, the early kink-type modulation phase of the positron filaments is strongly quenched. The saturation of the kink instability marks the onset of magnetic turbulence as observed at $\wi t \,=\, 750$. As we will further discuss in \S \ref{sec:coupling}, it is this microturbulence that controls the heating and coupling between leptons and ions in the long term evolution of the system.

\subsection{High multiplicity ($\M\,\gg\,1$)}
\label{sec:HighM}

In the high-multiplicity regime, the pair-ion CFI is the dominant electromagnetic mode, since the thermal spread of the pairs largely exceeds the relative drift between them, as discussed in \S\ref{sec:kinetic}. To verify this and to study the overall evolution of the system, we have performed a series of PIC simulations with multiplicities $\M\,=\,10, 25, 50$ and mass ratios $\ri\,=\,25, 100$. For these simulations, we used different values the radiation force ranging from $\Frad\,=\,2\times10^{-5}$ to $\Frad\,=\,10^{-3}$.

We have found that while indeed the pair-ion CFI is the dominant mode in the long-term development of magnetic turbulence, the choice of simulation setup can affect the details of the early time growth of instabilities in the high-multiplicity regime. This is because the longitudinal radiation force leads to a temperature anisotropy through adiabatic compression of the pair plasma, $P\,\propto\,n^{\Gamma_{\rm ad}}$. As large enough thermal anisotropies develop~\citep{Davidson_1972}, the system then becomes unstable to the Weibel instability ~\citep{Weibel_1959}. 
This is not the case in the low-multiplicity regime, where the rapid growth of the CFI produces sufficient microturbulence early on to prevent significant thermal anisotropies to develop. This raises the question of the most appropriate choice of frame for a periodic simulation configuration of the high-multiplicity regime, as depending on whether the plasma flow is decelerated or accelerated in this frame, the Weibel modes will be transverse or longitudinal to the flow. Note, however, that in the high-multiplicity regime the growth and saturation level of the Weibel instability is solely dependent on the pair plasma density and temperature due to the slow growth of the pair-ion CFI. We have modeled the system in different frames and found that the choice of initial configuration does not significantly affect the long-term dynamics of the system as the ions remain largely unmagnetized and the early Weibel-mediated turbulence itself quenches the thermal pressure anisotropy of the pairs. 
In Fig.~\ref{fig:M1_PIC}(c) we show results for $\M\,=\,10$, $\mime\,=\,25$, and $\Frad\,=\,9.5\times10^{-5}$ using a configuration where the plasma is initially drifting with four-velocity $u_0\,=\,-2$. The growth of the Weibel instability is seen at $\wi t\,\simeq\,100$. By $\wi t\,\simeq\,200$, the adiabatic pressure anisotropy is strongly reduced and the subsequent pair-ion instability ensues. 
In the low multiplicity regime, we observed that filament disruption is associated with kink unstable modes. However, in the high-multiplicity regime we see that the onset of microturbulence is dominated by the transition to cavity modes~\citep{Ruyer_2015a,Naseri_2018,Peterson_2021} [see inset (b) in Fig.~\ref{fig:F_CFI}]. These modes are associated with the transverse magnetic pressure in the filaments, which expel the background plasma causing cavitation of the density and distortion of the filaments.
In all the high-multiplicity cases simulated, we observe very similar dynamics. The pair-ion CFI consistently dominates the long-term evolution of the system and drives microturbulence during its nonlinear phase, similarly to what was observed in the low-multiplicity regime. We now turn our attention to the dynamics of the different species in this CFI-mediated turbulence.

\section{Inter-species coupling and heating rate}
\label{sec:coupling}

In the previous section, we discussed the dominant emerging linear microinstabilities growing in the precursor of a RRMS. As these instabilities grow, they affect the dynamics of the bulk plasma through the coupling of the different species with the electromagnetic microturbulence. In this section, we characterize the slowdown and heating of the plasma species via pitch angle scattering in the CFI-mediated microturbulence using semi-analytical and numerical modeling. 

We first develop the tenets of a reduced model that accounts for the scattering in pitch angle of the different species in the microturbulence. In \S \ref{sec:saturation}, we have seen that the scattering center frame can be estimated from the relative drifts between the ions and positrons. As the CFI is a magnetic instability and due to the acceleration of the leptons by the radiation drag force, the problem reduces to the pitch angle scattering of the particles in the noninertial frame of the quasi-magnetostatic microturbulence. Following a similar approach by ~\cite{Lemoine_2019_II}, originally based on~\cite{Webb_1989}, we model the dynamics of the plasma in mixed coordinates. We express the space coordinates in the shock front frame in which the system is stationary, and momentum coordinates in the scattering center frame in which momentum evolution is described as pitch-angle scattering.
In its general form, the relativistic transport equation of the distribution $f_s$ for the species $s$ is derived in the Appendix of~\cite{Lemoine_2019_II}. Including the radiation force $\Frad$ together with the electrostatic field in the stationary regime, the transport equation reads: 
\begin{eqnarray}
    & \gsc c \partial_x f_s - c \partial_x(\usc) p^t_{\rm sc}  \partial_{p^x_{|\rm sc}} f_s + \frac{{\Frad} + q_s E_x}{p^x_{|\rm sc}/p^t_{|\rm sc} + \bsc }   \partial_{p^x_{|\rm sc}} f_s \nonumber\\
    &\,=\,\frac{1}{2} \frac{1}{p^x_{|\rm sc}/p^t_{|\rm sc} + \bsc } \partial_{\mu_{|\rm sc}} \nu_{|\rm sc} \left( 1 - \mu_{|\rm sc}^2 \right) \partial_{\mu_{|\rm sc}}f \,,
\label{eq:transport}
\end{eqnarray}
where $p^t$ and $p^x$ are the respective time and longitudinal components of the four-momentum, and $\nu_{|\rm sc}$, $\mu_{|\rm sc}\,=\,p^x_{|sc}/p_{|sc}$ the respective scattering frequency and pitch angle cosine in the scattering center frame. Here, $\usc\,=\,\bsc \gsc$ is the four-velocity of the scattering center frame in the shock front frame. 

We derive a Fokker-Planck equation by performing a Legendre expansion in pitch angle cosine, $f = f^0(x,p_{|\rm sc}) + f^1(x,p_{|\rm sc}) \mu_{|\rm sc}$, where $f^1/f^0$ mainly accounts for the bulk drift in the scattering center frame. The equation for $f^0$ then reads
\begin{eqnarray}
    & \bsc c\partial_x f^0 - \frac{1}{3 \gsc} c\partial_x(\usc) p_{\rm |sc} \partial_{p_{\rm |sc}} f^0 \nonumber\\ 
    &\qquad\qquad\qquad\qquad- \frac{1}{p_{\rm |sc}^2} \partial_{p_{\rm |sc}} \left( p_{\rm |sc}^2 D_{p_{\rm |sc} p_{\rm |sc}} \partial_{p_{\rm |sc}}f^0\right) \,=\,0\,, \label{eq:Fokker-Planck}
\end{eqnarray}
to leading order in ${\Frad}$, $E_x$, and $\partial_x(\usc)$. Equation~\eqref{eq:Fokker-Planck} is identical to the Fokker-Planck equation describing the dynamics of the background pair plasma in relativistic Weibel-mediated shock waves~\citep{Lemoine_2019_PRL,Lemoine_2019_II} but with a modified diffusion coefficient accounting for the coupled effect of radiation and electrostatic forces (here in the limit $\gsc \gg 1$):
\begin{eqnarray}
    D_{p_{\rm |sc} p_{\rm |sc}} \,&=&\, \frac{p_{\rm |sc}^2}{3 \gsc \nu_{|\rm sc}} \biggl[ {\Frad}_{,s} + q_s E_x  \nonumber\\
    &&\qquad  - p^t_{|\rm sc} \left( 1 - \frac{p_{|\rm sc}^2}{3 p^{t 2}_{|\rm sc}} \right) c\partial_x(\gsc) \biggr]^2 \,. \label{eq:Dpp}
\end{eqnarray}
While the second term in Eq.~\eqref{eq:Fokker-Planck} corresponds to an adiabatic compression of the plasma, the third one describes a Joule-type heating of the particles in the electromagnetic microturbulence. The diffusion coefficient clearly indicates the different sources of friction via the relative drift imposed by the deceleration of the scattering center frame, $c\partial_x(\gsc)$, the electrostatic field, $E_x$, and the radiation force, $\Frad$. 

\begin{figure}
	\begin{center}
		\includegraphics[width=0.9
		\columnwidth]{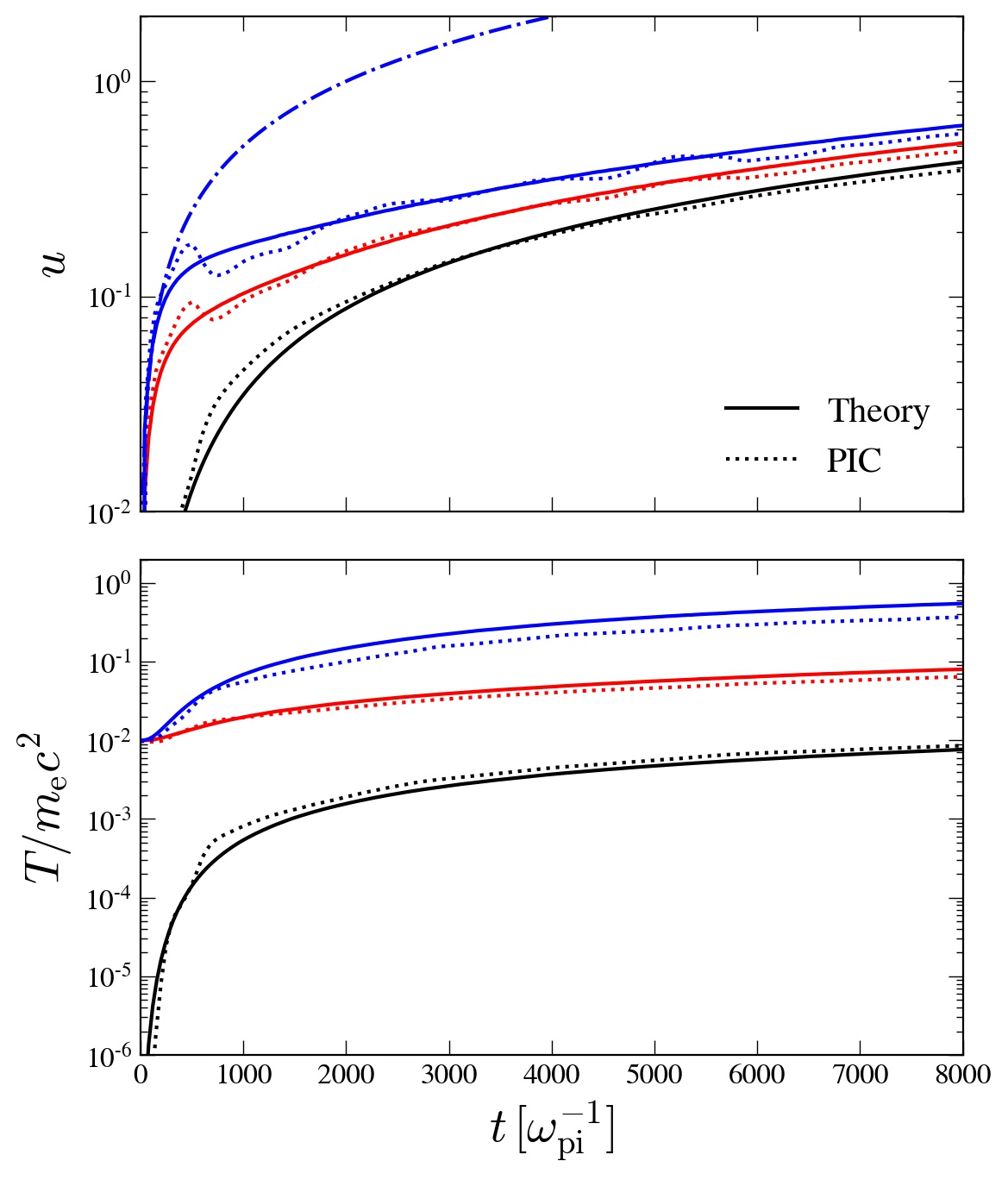}\\
		\caption{Evolution of the bulk velocity (top) and temperature (bottom) of the ions (black), electrons (red) and positrons (blue), obtained from the reduced MC model in non-inertial frame and comparison with a PIC simulation with $\Frad\,=\,2\times10^{-5}$, $\M\,=\,1$, and $\ri\,=\,25$ (dotted lines). The dot-dashed shows the positron trajectory without microturbulence. We assume energy independent scattering frequency for the particles, with $\nu_{\rm\pm|sc}=10^{-3}\wi$ for the leptons and $\nu_{\rm i|sc}=2\times10^{-4}\wi$ for the ions.
        }
		\label{fig:MC}
	\end{center}
\end{figure}

Note that the radiation force acts exclusively on the leptons. In that sense, the heating and deceleration of the ions is analogous to the dynamics of ions in unmagnetized collisionless shock waves. Another interesting feature, coming directly from the diffusion coefficient in Eq.~\eqref{eq:Dpp}, is that in the low-multiplicity regime ($\M\,\simeq\,1$), the electron heating is relatively low compared to the positrons. This is because the net radiation and electrostatic force acting on the positrons is twice the drag force acting on the electrons, resulting in a reduced diffusion coefficient for the electrons. Indeed, a lower electron heating is observed in the PIC simulations (see Fig.~\ref{fig:MC}). 

The above linear Fokker-Planck equation is valid in the regime of slow deceleration with respect to the scattering time and weak anisotropy of the distribution. To solve the full dynamics of the plasma in such a reduced model, we use the equivalence between the transport equation~\eqref{eq:transport} and a stochastic differential formulation of the problem~\citep{Risken}. The slowdown and heating of the plasma is thus computed by solving an equivalent It\^o-type stochastic differential equation:
\begin{eqnarray}
    &&\Delta\mu_{|sc} = \sqrt{2 \nu_{|sc} \Delta t_{|sc}}\,\chi \label{eq:ito1}\,,\\
    &&\Delta p^x_{|\rm sc} = p_{| \rm sc} \Delta \mu_{|\rm sc} - \left( p^t_{|\rm sc} + \beta_{\rm sc} p^x_{|\rm sc} \right) \partial_{t} \left(u_{\rm sc}\right) \Delta t_{|\rm sc} \nonumber\\
    && \qquad \qquad + \Frad \Delta t_{|\rm sc} + q E_x \Delta t_{|\rm sc} \,,
    \label{eq:ito2}
\end{eqnarray}
where Eq.~\eqref{eq:ito1} stands for the stochastic scattering in pitch angle where $\chi\,\sim\,\mathcal{N}(0,1)$ is a normally distributed variable. The second, third and last terms in the right-hand side of Eq.~\eqref{eq:ito2} respectively correspond to the noninertial contribution in the equation of motion, the radiation force and the electric force. %The latter is self-consistently solved via the net current of each species in the initial rest frame of the species. 
Practically, at each time step, the electric force is computed via the net current of the distribution in the initial rest frame of the species. The particles are then pushed in the local scattering center frame using Eqs.~\eqref{eq:ito1}-\eqref{eq:ito2}. The current is then updated. For more details, see~\cite{Vanthieghem_2019}.  

This model relies on two free parameters: the velocity of the scattering center frame and the scattering frequency $\nu_{|\rm sc}$. We note that, to allow comparison with our PIC simulations, we assume a homogeneous plasma ($c\partial_x\,=\,0$ and $\partial_{t}\,\neq\,0$) in our transport equation given that the scattering center velocity depends on time. [While Eq.~\eqref{eq:transport} assumes a stationary shock ($c\partial_x\,\neq\,0$ and $\partial_{t}\,=\,0$)]. In Fig.~\ref{fig:MC}, we compare the temporal evolution of the fluid velocity and temperature of each species obtained via the Monte-Carlo (MC) approach with the self-consistent 2D PIC simulations. We find very good agreement between the two, demonstrating that the heating and compression of the leptons and ions is indeed determined by pitch-angle scattering in the magnetic microturbulence.

\begin{figure}
	\begin{center}
		\includegraphics[width=0.75
		\columnwidth]{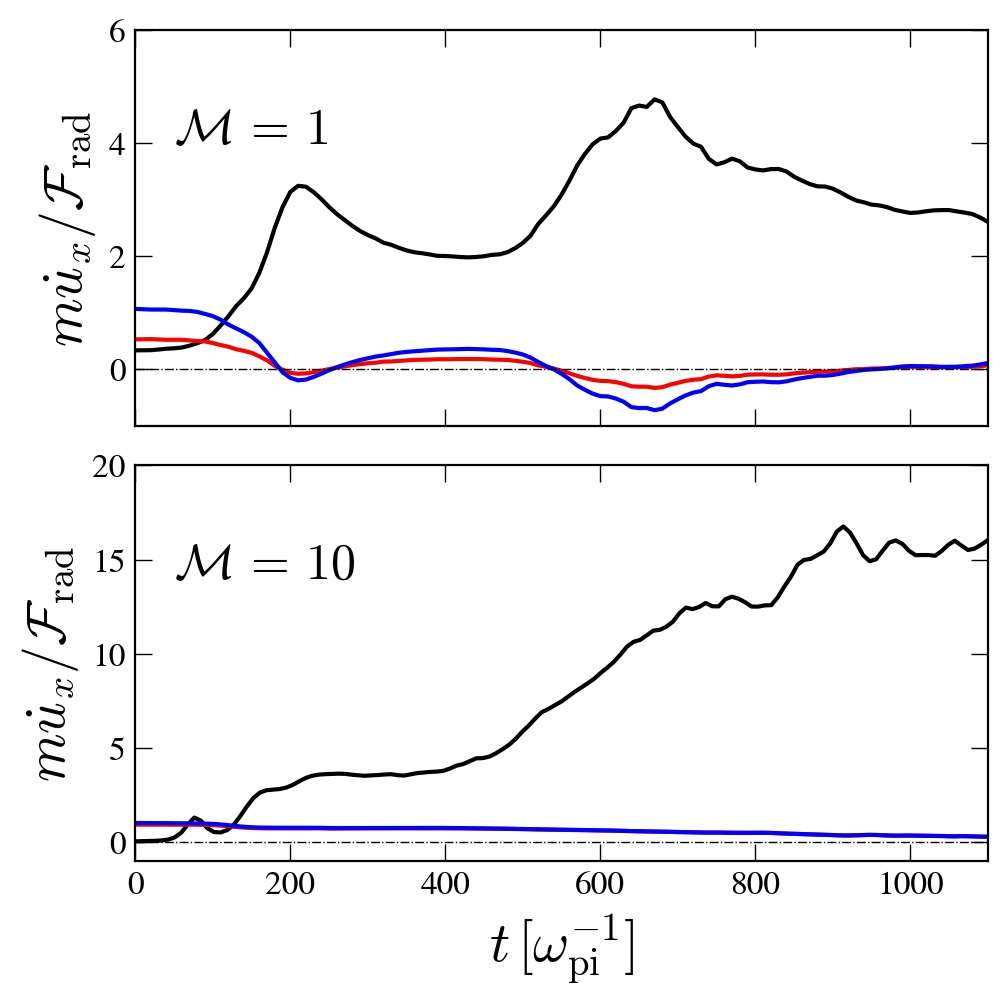}\\
		\caption{Bulk acceleration of the ions (black), positrons (blue), and electrons (red) for pair parameters $\M\,=\,1$, $\mime\,=\,100$ (top) and $\M\,=\,10$, $\mime\,=\,25$ (bottom) as obtained from PIC simulations discussed in \S \ref{sec:simulations} but for temperature $T_\pm\,=\,0.1 \me c^2$ of similar dynamics. The respective radiation forces are $\Frad\,=\,10^{-5}$ and $\Frad\,=\,9.4\times10^{-5}$.}
		\label{fig:coupling}
	\end{center}
\end{figure}

Figure~\ref{fig:coupling} displays the momentum transfer to the different species for various multiplicities obtained from the PIC simulations discussed in \S \ref{sec:simulations}. Before the onset of electromagnetic instabilities, species are purely accelerated by $\Frad$ and the induced electrostatic field. As $|\bp|\gtrsim0.1$ the pair-ion CFI develops and leads to the generation of microturbulence. Due to the coupling  with the microturbulence, leptons are progressively decelerated and most of the radiation force is efficiently transferred to the ions. 
The final level of the coupling (or relative velocity) reached between species depends on the ability of the drag force associated with the microturbulence to overcome the radiation force.

To characterize how the drag force due to microturbulence dependends on the multiplicity, mass ratio, and relative drift between the species in the absence of the external radiation force, we ran a series of 2D PIC simulations probing a parameter space encompassing $\mime\,=\,25-300$, $\M\,=\,5-100$, and $u_{0,\rm +}\,=\,3-10$ where $u_{0,+}$ is the initial relative drift between ions and positrons. (The initial electron drift velocity is determined by current neutrality.) The initial pair temperature is set to $T\,=\,0.1 \me c^2$. 

\begin{figure*}
	\begin{center}
		\includegraphics[width=0.8
		\textwidth]{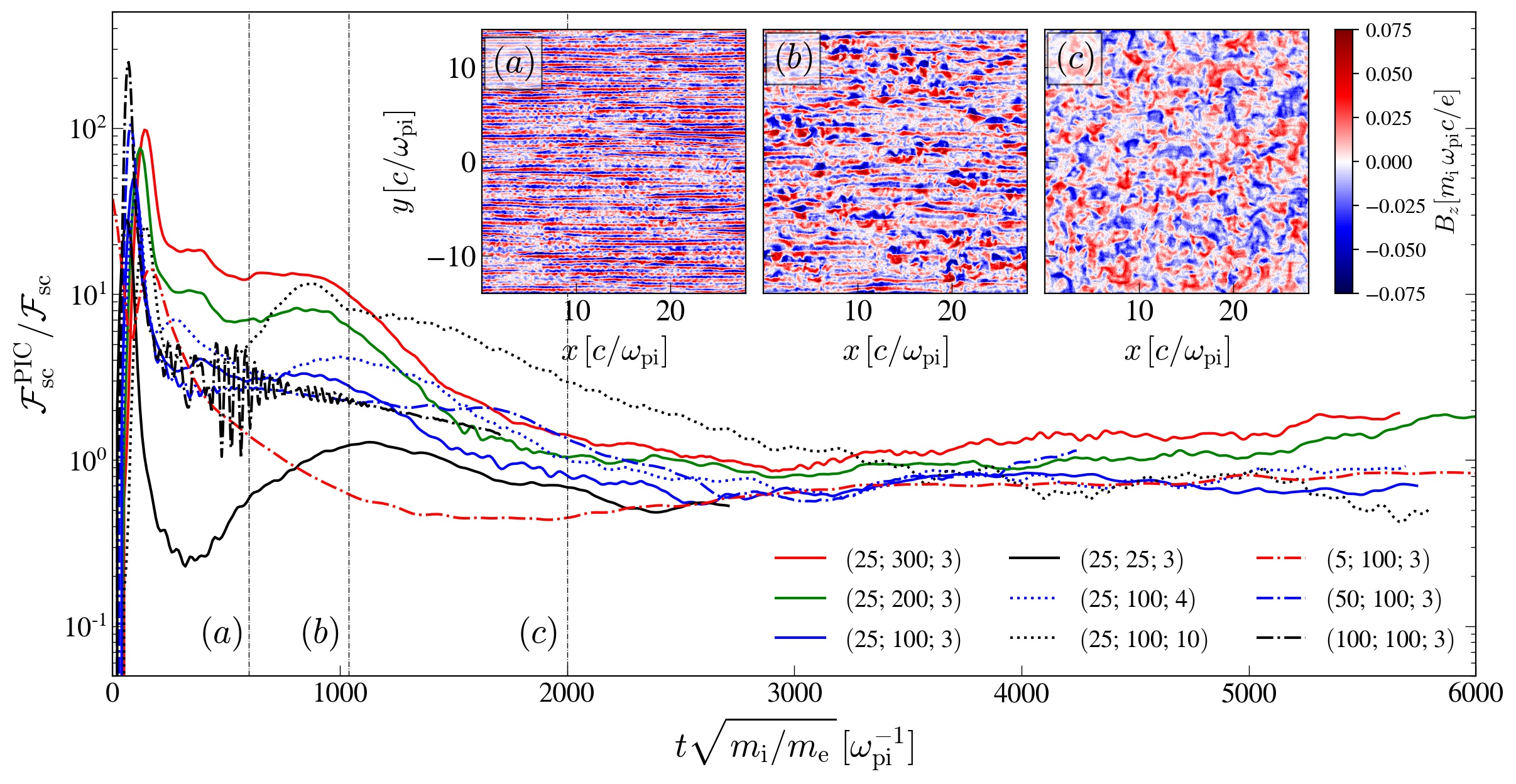}\\
		\caption{Evolution of the effective turbulence drag between the pairs and the ions, normalized by our estimate of $\Fw$ in Eq.~\eqref{eq:Fw}, for different parameters $(\mathcal{M},\mime,u_{\rm i})$. The inset shows representative profile of the (a) CFI , (b) nonlinear evolution, and (c) microturbulence for the simulation (25,100,4), taken at time corresponding to the respective vertical dot-dashed lines.  
        }
		\label{fig:F_CFI}
	\end{center}
\end{figure*}

All simulations show similar qualitative evolution. First, Buneman modes develop and contribute to modest ion heating and slightly reduce the relative drifts between species before saturating at low amplitude. These modes are then rapidly dominated by the pair-ion CFI [see inset (a) in Fig.~\ref{fig:F_CFI}]. 
As discussed in \S\ref{sec:saturation}, saturation  of the CFI modes occurs due to magnetic trapping of the ions in the ion scale filaments. While filaments and their disruption are mainly transient phenomena unable to fully couple the system, the resulting microturbulence [see inset (c) in Fig.~\ref{fig:F_CFI}] is sustained by the relative drift between species over long timescales, ensuring coupling. 

The leptons experience an effective drag force in the ion frame, due to the microturbulence, given by $\Fw\,=\,\me \frac{{\rm d}u_{\rm +}}{{\rm d} t } - \me \frac{{\rm d}u_{\rm i}}{{\rm d} t }\, \frac{\gamma_+}{\gamma_{\rm i}}$, where each term is expressed in an arbitrary inertial frame and the second term accounts for noninertial corrections. In the PIC simulations we observe that this drag force reaches a steady value, $\hatFw$, at the onset of microturbulence in all cases, as seen in Fig.~\ref{fig:F_CFI}. This drag force is sustained up to the point at which the relative drift between pairs and ions becomes sub-relativistic -- \emph{i.e.} the species are coupled. Based on the simulation results we obtain the following empirical scaling for the steady value
\begin{equation}\label{eq:Fw}
    \hatFw\,\sim\,10^{-1}\,\left(\frac{\me}{\mi}\right)^{3/2}  \, \M^{-1} \, u_{0,+}^{1/2}\,\mi \wi c\,.
\end{equation}

In the presence of both microtubulence and radiation force, the effective force acting on a lepton in the ion frame is $\Frad + \Fw$. In this case, the level of coupling between leptons and ions will be determined by $\hatFw/\Frad$. Following~\cite{granot2018}, in an infinite shock with upstream density of the order of $10^{15}\,{\rm cm}^{-3}$, the mean radiation force per particle, acting on a single lepton is of the order of $\Frad\,\sim\, 5\times10^{-10}\,\ginf^2/\gamma^2 \mi \wi c$ up to a prefactor of the order of unity, where $\ginf$ and $\gamma$ are the respective upstream and local Lorentz factors. For an hydrogen plasma, we thus obtain that the microturbulence drag force will couple the species when
\begin{eqnarray} \label{eq:cond_PIC}
    \frac{\hatFw}{\Frad} \sim 2.5\times10^3\,\frac{\gamma^2}{\ginf^2} \M^{-1}\,u_{0,+}^{1/2} \geq 1\,.
\end{eqnarray}

The drag force associated with the onset of the microinstabilities, when $|u_{0,+}|\,\sim\,0.1$, is defined as $\hatFwo \equiv \hatFw({u_{0,+}\,=\,0.1})$. If $\hatFwo/\Frad \geq 1$, the plasma species will be strongly coupled --- \emph{i.e.} coupling is quickly established and their relative drift remains subrelativistic. 
We can thus write a condition for strong coupling between the species as
\begin{eqnarray} \label{eq:coupling}
    \frac{\gamma}{\ginf} \geq \left(\frac{M}{10^3}\right)^{1/2}\,.
\end{eqnarray}
In the far precursor of RRMS, where the multiplicity remains sufficiently low, Eq.~\eqref{eq:coupling} is always satisfied and the species are thus expected to be strongly coupled by the microturbulence. However, in the relativistic regime and closer to the downstream, the local Lorentz factor drops to mildly relativistic values while the multiplicity increases up to values of the order of $\ginf \mime$. In this case, the coupling cannot be reached for $|\up|\,\sim\,0.1$ and the relative drift velocity between species will need to increase until strong enough turbulence is driven to ensure coupling, \emph{i.e.,} $\hatFw/\Frad\,\sim\,1$.

\begin{figure}
	\begin{center}
		\includegraphics[width=0.9
		\columnwidth]{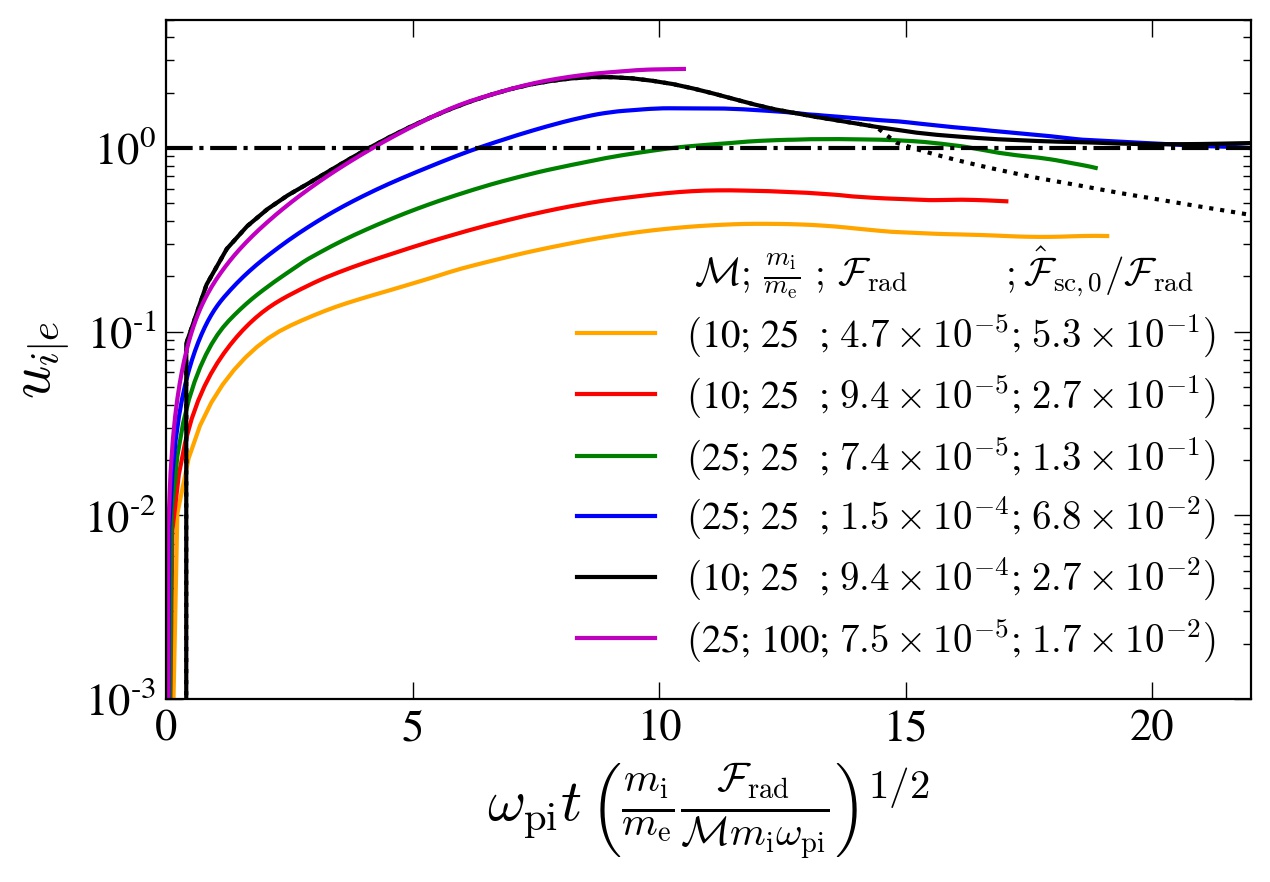}\\
		\caption{Drift velocity of the ion in the electron frame for PIC simulations with $\hatFwo/\Frad$ ranging from $1.7\times10^{-2}$ to $5.3\times10^{-1}$ for parameters $(\M,\,\mime,\,\Frad,\,\hatFwo/\Frad)$. For the dotted black line, the external radiation force is suppressed at $\omega_{\rm pi} t\,=\,300$.}
		\label{fig:var_rad}
	\end{center}
\end{figure}

To confirm how the level of coupling between species depends on the ratio $\hatFwo/\Frad$, we have performed a series of 2D PIC simulations in the high-multiplicity regime using the same configuration described in \S\ref{sec:HighM}.
The simulation results are summarized in Fig.~\ref{fig:var_rad} and show that indeed for $\hatFwo/\Frad\,\sim\,1$ strong coupling between species is quickly established and the relative drift remains subrelativistic. As $\hatFwo/\Frad$ decreases we see that the relative drift between pairs and ions increases, as expected. Interestingly, we find an asymptotic behavior for $\hatFwo/\Frad\,\ll\,1$, where the relative drift between pairs and ions saturates at $\up\,\sim\,1$.

We note that in reality as the species enter the downstream region the net radiation force experienced decreases, thus increasing $\hatFwo/\Frad$ and strongly coupling the species via the microturbulence. Indeed, as shown on Fig.~\ref{fig:var_rad}, when we repeat the PIC simulations with ${\hatFwo}/\Frad \ll 1$ and turn off the radiation force after the coupling with $u_+ \sim 1$ is reached, we observe that this residual relative drift quickly decreases and species are strongly coupled.

Based on the simulations results, we find that the coupling time due to the microturbulence is $\omega_\mathrm{pi}\,\tau_c \sim 10 [\M \me/(\mi \Frad)]^{1/2}$, which corresponds to a coupling length in the shock front frame $l_c \sim 6\times 10^4 [\M (10^{15} \mathrm{cm}^{-3}/n_i)]^{1/2}\,(\ginf/10)^{-1}$ cm. While large when compared to the plasma skin depth, this coupling length is much smaller than the RMS width which is found numerically to be $l_{RMS} \sim  10^{-3} \lambda_T \ginf^3 \approx 10^8 (10^{15} \mathrm{cm}^{-3}/n_i)\,(\ginf/10)^{2}$ cm \citep{ito2020}, where $\lambda_T = (\sigma_T n_i \gamma_\infty)^{-1}$ is the  Thomson length.

\section{Role of the ambient magnetic field}
\label{sec:instabilitymagnetized}

So far our analysis considered that initially the ambient plasma is unmagnetized. In the presence of a background magnetic field perpendicular to the shock front, the coupling between leptons and ions can be significantly stronger than in the unmagnetized case. While ions are decelerated electrostatically by the radiation force, the deceleration of magnetized leptons is reduced due to the restoring forces of electric fields and velocities growing along the shock front. Large magnetizations enhance this coupling such that the velocity spread between the lepton and ion fluids becomes small and can prevent the development of microinstabilities. In the following, we use the results from various fluid approximations to determine a limit on the background magnetization above which the instabilities considered in the previous sections will likely cease to develop.

First, we establish the relevant --- low frequency --- response of the plasma's lepton component. On time scales on which the radiation force does not significantly decelerate the ions, leptons behave like an isolated system. The characteristic frequencies of their oscillations (around the ions) can be determined by a suitable linear analysis of the dielectric properties of an electron-positron plasma in a constant magnetic field. Leptons oscillate with frequencies close to the plasma frequency, and also on a low frequency branch where
\begin{align}
  \omega_\pm&=\frac{m_i}{m_e}\frac{1}{\ginf^2}\frac{\sqrt{\sigma}}{1+2 \mathcal{M}}\sim 10^{-2}\left[\frac{\ginf}{10}\right]^{-2} \mathcal{M}^{-1}\left[\frac{\sigma}{10^{-6}}\right]^{1/2}.
    \label{eq:lowfreqbranch}
\end{align}
Here, $\ginf$ is the Lorentz factor of the upstream flow, and $\sigma=B^2/(4\pi  m_i n_i \ginf^2 c^2)$ is the magnetization far upstream.
The low frequency branch specified in Eq.~(\ref{eq:lowfreqbranch}) characterizes the time scales of the lepton-ion coupling. 

Second, we examine how lepton velocities scale with plasma and radiation characteristics. The velocity spread between the ion and pair fluids can be estimated using the following heuristic argument.
The radiation force ${\mathbfcal{F}}_{\rm rad}= -|{\mathbfcal{F}}_{\rm rad}|\hat{x}$ induces a transverse drift of electrons and positrons in opposite directions.
The longitudinal electric field generated inside the shock, $\mathbf{E}_\parallel = E_x\hat{x}$, induces a transverse drift in the
same direction. The net drift velocity of species $\alpha$ in the shock frame is 
\begin{equation}
\pmb{\beta}_{d,\alpha} = \left(\mathbf{E}_\parallel +\frac{1}{q_\alpha} {\mathbfcal{F}}_{\rm rad}\right)\times \frac{\mathbf{B}}{B^2},
\label{eq:drift1}
\end{equation}
where  $q_\alpha$ is the charge, and $\mathbf{B}= B_z \hat{z}$ is the local magnetic field inside the shock. 
Since the transverse electric current induced by the drift, $\mathbf{j}_{\perp} = e(n_+\gamma_+\pmb{\beta}_{d,+} - n_{-}\gamma_-\pmb{\beta}_{d,-})$, must
nearly vanish, we have:
\begin{equation}
{\cal M}\pmb{\beta}_{d,+} = ({\cal M}+1)\pmb{\beta}_{d,-}\;.
\label{eq:drift2}
\end{equation}
Defining $\mathbf{u}_{d,\alpha} = \gamma\pmb{\beta}_{d,\alpha}$, where the Lorentz factor $\gamma$ is roughly equal for all species when the coupling is strong, the solution to Eqs. (\ref{eq:drift1}) and (\ref{eq:drift2}) is
\begin{align}
\begin{split}
    \mathbf{u}_{d,+}+\mathbf{u}_{d,-} &= (1+2{\cal M})(\mathbf{u}_{d,+}-\mathbf{u}_{d,-})\\
&= 2(1+2{\cal M}) \frac{\gamma \mathrm{F}_{\rm rad}}{eB}\hat{y}\label{eq:transversedrift},
\end{split}\\
e\mathbf{E}_\parallel &= (1+2{\cal M})\mathbf{F}_{\rm rad}.
\label{eq:parallelE}
\end{align}
The transverse drift (Eq.~\ref{eq:transversedrift}) is comparable to the longitudinal one in the shock frame, as to say, it dictates the amplitude of the velocity spread along the direction of $\mathbfcal{F}_{\rm rad}$. Hereafter, we denote this spread (in the instantaneous ion frame) as the sum of deviations $\delta u$ from the ion velocity, and define $\xi^x = \delta u_+^x + \delta u_-^x$. For comparison to the theory laid out in the previous sections, a quantification in the ion frame is vital. In a semi-analytic analysis of a mildly magnetized three-fluid approximation with a constant radiation force, we recover the heuristic estimates of Eqs.~(\ref{eq:transversedrift}) and~(\ref{eq:parallelE}). Specifically, by an analysis of the corresponding differential operator, one finds a relation for the dominant longitudinal velocity spread:
\begin{align}
    \mathcal{A}\left[\xi^x_+\right]&=2\left(1+2\mathcal{M}\right)\frac{\mathcal{F}_{\rm rad}}{\sqrt{\sigma}}\label{eq:ampp}
\end{align}

Eq.~(\ref{eq:ampp}) can be used to derive an average velocity of the lepton fluid and we can establish a somewhat conservative limit for the instability growth. Motivated by a sharp decline of the instability growth rate in the hot case at low values of the streaming velocity, we assume the limit $\mathcal{A}\left[\xi^x_+\right]/2\gtrsim u_+$ or, in other terms,
\begin{align}
\begin{split}
        \sigma &\lesssim 4\mathcal{M}^2\frac{\mathcal{F}_{\rm rad}^2}{u_+^2}=4.0\times 10^{-6}\left[\frac{0.1}{u_+}\right]^2\left[\frac{\mathcal{F}_{\rm rad}}{10^{-7}}\right]^2\left[\frac{\mathcal{M}}{10^3}\right]^2,
    \label{eq:sigmaconservative}
\end{split}
\end{align}
noting that in RRMS the multiplicity well inside the shock exceeds the mass ratio, ${\cal M}\gtrsim m_i/m_e$.

Finally, we scrutinize the conservative estimate in Eq.~(\ref{eq:sigmaconservative}) by probing the underlying assumption of $u_+\approx 0.1 $ against (semi-)analytic models of the growth rate $\Gamma$. For $ T_\pm\,=\,\me c^2$, and $\mathcal{M}=100$, we find $\Gamma\approx 0.1$ (cf. Fig.~\ref{fig:G_u_th}). Reversing the arguments from above we can, thus, demand $\omega_\pm\lesssim \Gamma/10^3\approx 10^{-4}$, as to say in other words $\sigma\lesssim 10^{-4}\left[\mathcal{M}^2/10^6\right]$. A self-consistent modeling of the interplay between the deceleration and multiplicity $\mathcal{M}$, and a comprehensive instability analysis for a magnetized environment will be the focus of a subsequent work.

Estimating where magnetic fields play a role in RRMS in nature is difficult, mostly because the value of $\sigma$ in astrophysical RRMS environments varies from one system to another, and is not well constrained in most systems. One example is GRB jets where $\sigma$ is uncertain, though most likely non-negligible (note, however, that at least some shocks in GRB jets are photon rich and thus contain no pairs). Another example
where RRMS play a dominant role is during the breakout from a stellar envelope. Even though only few measurements of the magnetic field strength on the surface of massive stars have been reported, a robust upper limit on $\sigma$ at the breakout radius of an RRMS can be placed: since the magnetic energy density at the stellar surface must be much lower than the gravitational energy density we obtain $\sigma \ll 10^{-6} (R/10^{11} {\rm cm})^{-1} (M/5M_\odot) (\ginf/3)^{-2}$, where $R$ and $M$ are the stellar radius and mass respectively. Thus, we anticipate that in some systems magnetic fields may considerably alter our results and in many others they can be neglected.

\section{Summary}
\label{sec:conclusions}

This paper addresses a fundamental question in the theory of relativistic radiation mediated shock waves: What is the mechanism that couples the different plasma constituents (ions, electrons and positrons), and how does it affect the shock thermodynamics and emission?  Since the presence of positrons prevents electrostatic coupling, plasma instabilities appear as critical to couple ions and pairs inside the shock. 
To that end, we carried out a comprehensive stability analysis of kinetic effects in such shocks.  Because the dramatic scale separation between radiation and kinetic scales renders the construction of a global shock model impractical, we based our analysis on a simplified model that treats the radiation drag as a fixed external force acting solely on the leptons, ignores pair creation processes, and invokes uniformity. 
Such conditions are anticipated, approximately, on scales over which the microturbulence develops, which are much shorter than the RMS width. 

Our strategy was to perform a linear stability analysis in order to elucidate the instability criteria (specifically, velocity separation threshold, and its dependence on temperature and pair multiplicity), and to compare the results 
with PIC simulations that compute the nonlinear evolution of the instability and its saturation.

Our main conclusion is that in weakly magnetized plasmas the different species couple via the electromagnetic turbulence excited by a current filamentation instability. The plasma becomes unstable when the relative drift velocity between pairs and ions satisfies $|\bp|\,\gtrsim\,0.1$. The ensuing microturbulence on plasma kinetic scales governs the momentum transfer between pairs and ions, and leads to nonadiabatic heating of the particles. For the typical radiation profile expected in RRMS, for low to moderate pair multiplicities ($\M \ll 10^3$), strong coupling between species is quickly established and the relative drift remains subrelativistic. For high multiplicities ($\M \gtrsim 10^3$), the relative four-velocity between species increases but, as observed in PIC simulations, is bounded by $u_+ \sim 1$. 

The generation of magnetic turbulence in RRMS can have important implications beyond the coupling of the different species. In particular, it can potentially convert a fraction of the dissipated energy to nonthermal particles with power-law distributions. Both the increased nonadiabatic heating found in our study and the possible formation of power-law energy spectra can impact the shock breakout emission. Finally, the residual relative drift that is established between pairs and ions in high multiplicity regions should motivate further kinetic studies that can capture its role on the shock structure, namely the potential formation of collisionless subshocks.

\section{Data Availability}
The data underlying this article will be shared on reasonable request to the corresponding author.

\section{Acknowledgments}
AV and FF acknowledge support by the U.S. DOE Early Career Research Program under FWP 100331. AL and EN acknowledge support by the Israel Science Foundation grant 1114/17. AP and JM acknowledge support by the National Science Foundation under Grant No. AST-1909458. Research at the Flatiron Institute is supported by the Simons Foundation. The authors acknowledge the OSIRIS Consortium, consisting of UCLA and IST (Portugal) for the use of the OSIRIS 4.0 framework. Simulations were run on CORI at the National Energy Research Scientific Computing Center (NERSC) through ALCC award.

%%%%%%%%%%%%%%%%%%%% REFERENCES %%%%%%%%%%%%%%%%%%

\bibliographystyle{mnras}
\bibliography{literature} 

%%%%%%%%%%%%%%%%%%%%%%%%%%%%%%%%%%%%%%%%%%%%%%%%%%

%%%%%%%%%%%%%%%%% APPENDICES %%%%%%%%%%%%%%%%%%%%%

\appendix

\section{Kinetic estimate for a warm pair plasma}
\label{sec:appA}

Here, we evaluate the transverse modes growing in a warm pair plasma of multiplicity $\M$ drifting at subrelativistic to mildly relativistic speed in a cold ion background. We briefly recall the tenets of the formalism~\cite{Silva_2002}. The well established linear dispersion relation fulfilled by the purely longitudinal and transverse modes are respectively
\begin{eqnarray}
    \epsilon_{xx} \left( \epsilon_{yy} \omega^2 - c^2 k_x^2 \right)- \epsilon_{xy}^2   \,=\, 0\,\label{eq:disp_rel_kin_cold}\\
    \epsilon_{yy} \left( \epsilon_{xx} \omega^2 - c^2 k_y^2 \right)- \epsilon_{xy}^2   \,=\, 0\,,\label{eq:disp_rel_kin}
\end{eqnarray}
where $k_x$ and $k_y$ are the respective longitudinal and transverse wavenumbers. The non-vanishing off-diagonal components of the dielectric tensor $\epsilon_{ij}$ accounts for the system asymmetry [$\bp\,=\,(1+1/\M )\be$]. In a fully relativistic framework, $\epsilon_{ij}$ is obtained from the following expression:
\begin{equation}
    \epsilon_{ij} = \delta_{ij} + \sum_s \frac{\omega_{ps}^2}{\omega^2} \int \frac{ {\rm d}^3{\mathbf{p}}}{\gamma} \left( p_i \nabla_{p_j}  + \beta_i p_j \frac{k\cdot\nabla_{\mathbf{p}}}{ \omega - c {\mathbf{k}} \cdot \mathbf{\beta}}\right) f_s\,,
\end{equation}
for the species $s$. Here we focus on the transverse case ($k\,=\,k_y$) with cold ions and warm electrons. As we check \emph{a posteriori}, transverse modes start to grow for subrelativistic to mildly relativistic speeds. Therefore, we here approach the problem using a classical drifting 2D Maxwellian distribution for the pairs:
\begin{equation}
    f_s = \frac{n_s m_e c^2}{2 \pi T} e^{-\frac{m_e c^2}{2 T} \left[ \left(p_x - p_d\right)^2 + p_y^2 \right]}\,,
\end{equation}
where $p_d\,=\,\beta_d$ is the bulk drift speed; and the ions are modeled by a cold distribution at rest. Searching for purely transverse and growing modes, it is useful to introduce $z\,=\,\sqrt{\me/2T}\,\Gamma/k$. One can then compute the relevant components of the dielectric tensor:
\begin{align}
    \epsilon_{xx}&\,=\,1 - \frac{\wi^2}{\omega^2} - \frac{\wi^2}{\omega^2} \M \rif  \left\{ 1 - \left(\frac{\bp^2}{T/\me c^2}  + 1 \right)  \left[ 1 - z \tilde{\mathcal{Z}}(z) \right] \right\}  \nonumber \\
    &\qquad - \frac{\wi^2}{\omega^2} \left(\M +1 \right) \rif  \left\{ 1 - \left(\frac{\be^2}{T/\me c^2} + 1 \right)  \left[ 1 - z \tilde{\mathcal{Z}}(z) \right]\right\}\\
    \epsilon_{yy}&\,=\,1 - \frac{\wi^2}{\omega^2} - \frac{\wi^2}{\omega^2} \left(2\M+1\right) \rif  2 z^2 \left[1 - z \tilde{\mathcal{Z}}(z) \right]  \\
    \epsilon_{xy}&\,=\, - 2 i\,\bp\,\sqrt{\frac{2}{T/\me c^2}}\,z\,\left[ 1 - z \tilde{\mathcal{Z}}(z) \right]
\end{align}
where $\tilde{\mathcal{Z}}(z)\,=\,-i \mathcal{Z}(iz)\,=\,\sqrt{\pi} \exp(z^2) \left[ 1 - {\rm erf}(z) \right]$ with $\mathcal{Z}$ the plasma dispersion function. In the regime of interest where the relative drift speed between ions and positrons $|\bp|\sim0.1$ and the pair are warm $T\simeq m_e c^2$, we have that $z\ll1$ and we thus expand the dispersion function in this limit.
For a warm electron-positron plasma of multiplicity $\M \gg 1$, the dispersion relation reads:
\begin{eqnarray}
    && 4 z^2  \left[ \frac{\left(c k/\wi \right)^2}{\ri \mathcal{M}} + 2\sqrt{\pi} z \right] \nonumber \\
    &&+ \frac{1}{\ri \mathcal{M}}  \left[ \frac{ \left(c k/\wi\right)^2}{\ri \mathcal{M}} + 2 \sqrt{ \pi} z - 2 \frac{\bp^2}{T/\me c^2} \right]\,=\,0\,.\label{eq:disp_Max}
\end{eqnarray}
The above Eq.~\eqref{eq:disp_Max} has been obtained in the limit $z\ll1$, which holds when the relative drift between ions and pairs is weak and the thermal spread is high.

%%%%%%%%%%%%%%%%%%%%%%%%%%%%%%%%%%%%%%%%%%%%%%%%%%
% Don't change these lines
\bsp	% typesetting comment
\label{lastpage}
\end{document}